\RequirePackage{fix-cm}
\documentclass[twocolumn]{svjour3}          
\smartqed  
\usepackage{graphicx}
\usepackage{physics}
\usepackage{subfigure}
\usepackage{float}
\usepackage{hyperref} 
\usepackage{mathtools}
\usepackage[authoryear, round]{natbib}
\usepackage{multirow}
\usepackage{balance}
\usepackage{xurl}
\usepackage{algorithm} 
\usepackage{algpseudocode} 
\newcommand{\eprint}[2][]{arXiv:\href{https://arxiv.org/abs/#2}{\nolinkurl{#2}}} 
\newcommand{\doi}[2][]{DOI:\href{https://doi.org/#2}{\nolinkurl{#2}}} 
\journalname{Quantum Machine Intelligence}
\begin{document}

\title{Hybrid Quantum Classical Graph Neural Networks for Particle Track Reconstruction
}


\author{
Cenk T\"uys\"uz~$^{1}$ 
    \and Carla Rieger~$^{2}$ 
    \and Kristiane Novotny~$^{3}$ 
    \\ Bilge Demirk\"oz~$^{1}$ 
    \and Daniel Dobos~$^{3,4}$ 
    \and Karolos Potamianos~$^{3,5}$ 
    \\ Sofia Vallecorsa~$^{6}$ 
    \and Jean-Roch Vlimant~$^{7}$ 
    \and Richard Forster~$^{3}$ 
}


\institute{
Corresponding author: Cenk T\"uys\"uz\\
\email{cenk.tuysuz@cern.ch}\\
$^{1}$~Department of Physics, Middle East Technical University, Ankara, Turkey
\\$^{2}$~Department of Physics, ETH Z\"urich, Z\"urich, Switzerland
\\$^{3}$~gluoNNet, Geneva, Switzerland
\\$^{4}$~Lancaster University, Lancaster, UK
\\$^{5}$~University of Oxford, Oxford, UK
\\$^{6}$~CERN, Geneva, Switzerland,
\\$^{7}$~California Institute of Technology, Pasadena, California, USA
}

\date{Received: date / Accepted: date}

\maketitle

\begin{abstract}
The Large Hadron Collider (LHC) at the European Organisation for Nuclear Research (CERN) will be upgraded to further increase the instantaneous rate of particle collisions (luminosity) and become the High Luminosity LHC (HL-LHC). This increase in luminosity will significantly increase the number of particles interacting with the detector. The interaction of particles with a detector is referred to as "hit". The HL-LHC will yield many more detector hits, which will pose a combinatorial challenge by using reconstruction algorithms to determine particle trajectories from those hits. This work explores the possibility of converting a novel Graph Neural Network model, that can optimally take into account the sparse nature of the tracking detector data and their complex geometry, to a Hybrid Quantum-Classical Graph Neural Network that benefits from using Variational Quantum layers. We show that this hybrid model can perform similar to the classical approach. Also, we explore Parametrized Quantum Circuits (PQC) with different expressibility and entangling capacities, and compare their training performance in order to quantify the expected benefits. These results can be used to build a future road map to further develop circuit based Hybrid Quantum-Classical Graph Neural Networks. 
\end{abstract}

\section{Introduction}
\label{intro}
Particle accelerator experiments aim to understand the nature of particles by colliding groups of particles at high energies and try to observe creation of particles and their decays, e.g. to validate theories. The Large Hadron Collider~(LHC) at the European Organisation for Nuclear Research~(CERN) provides proton-proton collisions to four main experiments as well as other small experiments and fixed-target experiments. In order to achieve a high sensitivity, these experiments use advanced software and hardware.

In addition, these experiments will require very fast processing units as the time between two consecutive collisions is very short (reaching up to 1 MHz for ATLAS and CMS according to \cite{atlas-report}, \cite{cms-report}) and \cite{lhc-computing}. A big data storage and processing problem arises, when the fast data acquisition is combined with sensitive hardware. A total disk and tape spaces of 990 PetaBytes and around 550 thousand CPU cores were pledged to LHC experiments in 2017 according to a report by CERN Computing Resources Scrutiny Group (CRSG) (\cite{computing-report}).

Currently, the LHC is going through an upgrade period to increase the number of particles in the beam (i.e. luminosity) (\cite{ref-hilumi}). Therefore, the future High Luminosity LHC (HL-LHC) experiments will require much faster electronics and software to process the increased rate of collisions. 

In particle track reconstruction problem, the aim is to identify the trajectory of particles using the measurements of the tracking detectors. Accelerated particles interact/collide near the origin of the coordinate system of detectors. Products of these interactions travel outwards from the origin. Charged particles bend in a direction depending on their electric charge. When these particles pass through the detectors, they create signals called as hits. Particle track reconstruction aims to connect hits belonging to the same particles to assign a trajectory.

The efficient reconstruction of particle tracks is one of the most important challenges in the HL-LHC upgrade. Although there are novel algorithms (\cite{atlas-fast-track,cms-ctf-2}) available that are able to handle the current rate of collisions, they suffer from higher collision rates as they scale worse than quadratically (e.g. $\mathcal{O}(n^6)$ (\cite{Magano2021}). 

\begin{figure}[!ht]
    \centering
    \includegraphics[width=\linewidth]{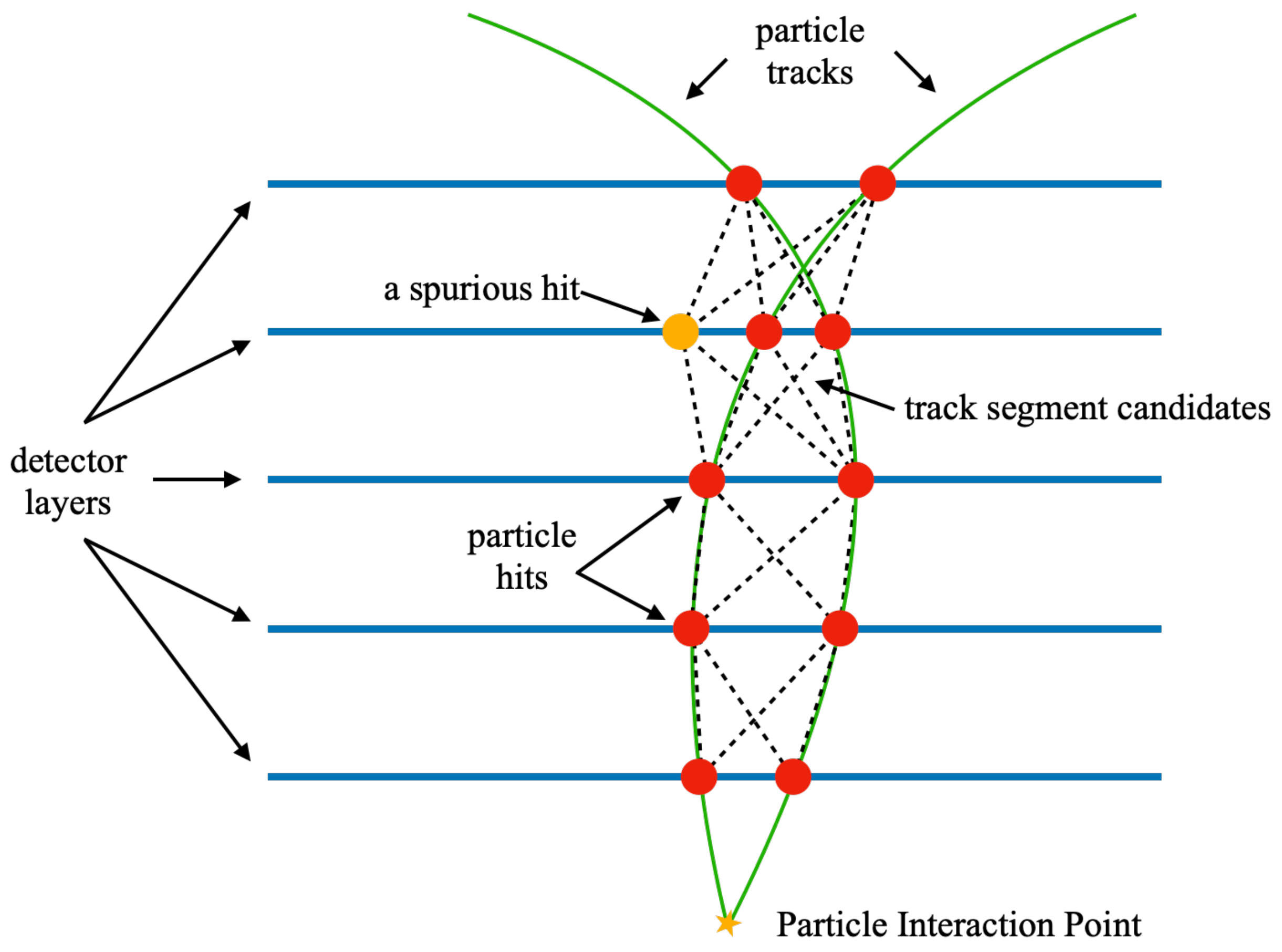}
    \caption{Drawing of particle track reconstruction problem. Particles interact near the origin of the coordinate system. Products of these interactions travel outwards from the origin. Charged particles bend in a direction depending on their electric charge. When these charged particles pass through the detectors, they create signals called as hits. Particle track reconstruction aims to connect hits belonging to the same particles.}
    \label{fig:my_label}
\end{figure}

Recent developments in Quantum Computing (QC) allowed scientists to look at computational problems from a new perspective. There is a great effort to make use of these new tools provided by QC to gain high speed-ups for many computational tasks in High Energy Physics (\cite{qml-hep}). There are many problems investigated. This include but not limited to physics analysis at LHC using kernel~(\cite{wu_application_2021, heredge_quantum_2021}) and variational methods~(\cite{wu_application_2021-1, terashi_event_2021}), simulating parton showers~(\cite{jang_quantum_2021}) and imitating calorimeter outputs using Quantum Generative Adversarial Networks~(\cite{chang_dual-parameterized_2021}).  

Researchers have been investigating QC tools for a computational advantage for the particle track reconstruction problem, since it also suffers from scaling. While there are several attempts using adiabatic QC (\cite{qalg2,qalg3}), Quantum Associative Memory~\cite{shapoval_quantum_2019} or Quantum search routines (\cite{Magano2021}), this work focuses on hybrid variational methods. 

In this work, we aim to give a complete overview on our developments, where we investigated the use of a Hybrid Quantum-Classical Graph Neural Network (QGNN) approach to solve the particle track reconstruction problem (\cite{tuysuz-1,tuysuz-2, tuysuz-3}) that has been trained on the publicly available TrackML Challenge dataset (\cite{trackml, trackml-2}). We present an analysis of several well-performing Quantum Circuits and give a comparison with its classical equivalent, HEP.TrkX (\cite{HepTrkX}), on which our approach is based on. 

The rest of the paper is organized as follows. Details of the dataset and pre-processing methods are given in Section~\ref{dataset}. The QGNN model is explained in detail in Section~\ref{hgnn}. Results and comparisons with novel methods are given in Section~\ref{results}, along with a discussion of the findings. Finally, our summary and comments on possible improvements are presented in Section~\ref{conclusion}.

\section{The Dataset and Pre-processing}
\label{dataset}

\begin{figure}[!ht]
    \centering
    \includegraphics[width=\linewidth]{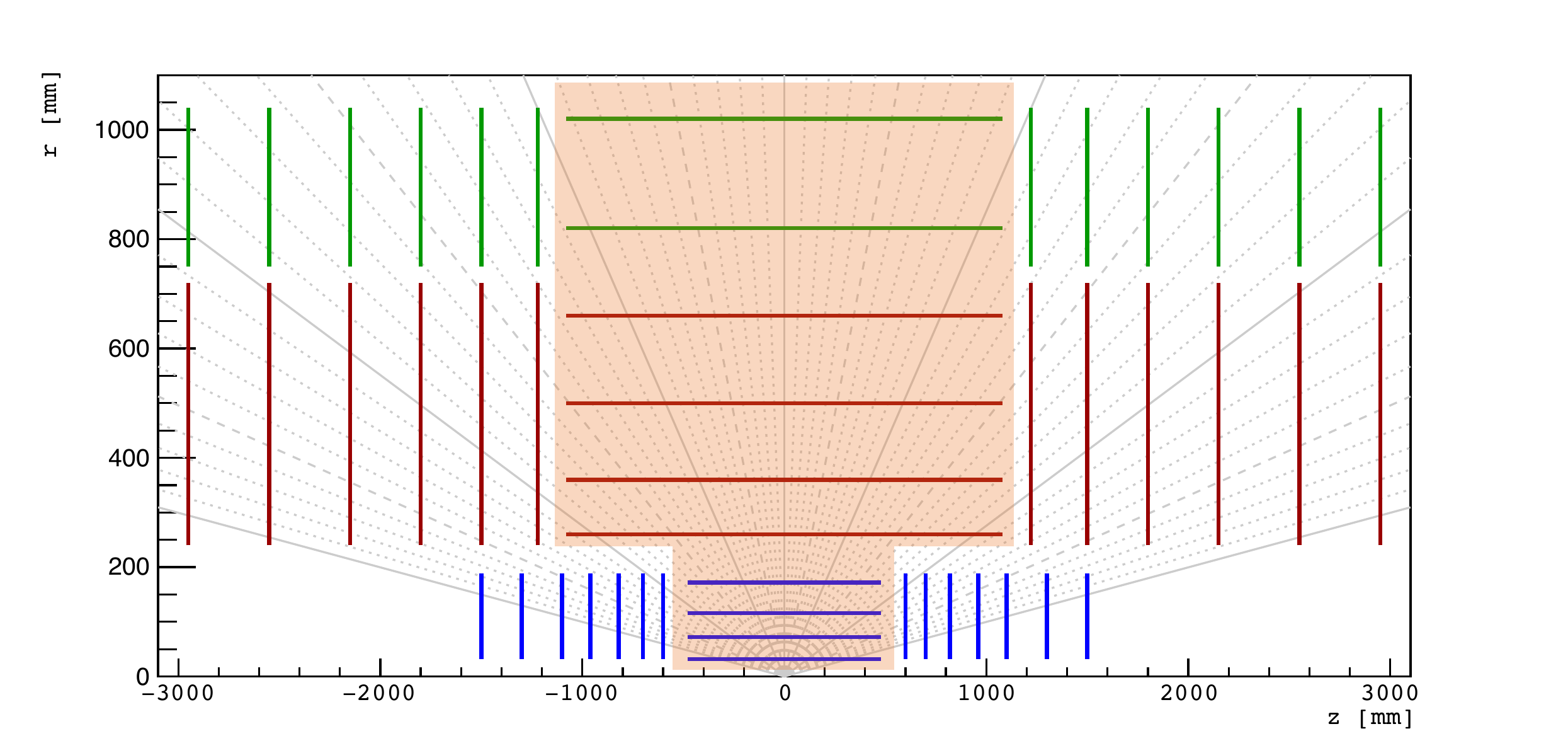}
    \caption{TrackML detector geometry projected to 2 dimensions ($r$, $z$). The region highlighted in orange indicates the detector layers used in this work. Drawing is adapted from \cite{trackml}.}
    \label{fig:trackml}
\end{figure}

The publicly available TrackML Challenge dataset provides 10000 events  to emulate the HL-LHC conditions (\cite{trackml}). It has become a benchmark dataset for researchers after the conclusion of the challenge~\cite{trackml-2} and allows comparisons across different methods. The simulated tracking detector geometry of the dataset is that of a general purpose collider experiment. The schema of this geometry in 2 cylindrical coordinates (r,z) can be seen in Fig.~\ref{fig:trackml}. Horizontal layers in the center of the detector represent a barrel-shaped geometry, while vertical layers represent a disk-shaped geometry and are generally referred to as end-cap layers. 

In many Quantum Machine Learning applications, it is very hard to work with large datasets due to restrictions on simulation times. A pre-processing step is necessary, which reduces the amount of samples and prepares the data format for the model.

The pre-processing procedure starts by selecting the first 100 events from the dataset. Although it would be ideal to use all events of the dataset, computation time restrictions of QC simulation limited us to use only a portion of the dataset.  Then, particle hits are restricted to  the barrel region of the detector, which is the region highlighted in Fig.~\ref{fig:trackml}. This limits the number of tracks and the ambiguity in identifying the particle trajectories. In addition, a cut in the transverse momentum of the particle  is applied to further reduce the number tracks.

After reducing the number of tracks to reasonable numbers, the next step is to create graphs out of the remaining particle hits. Particle hits become the nodes and the track segment candidates will be defined as edges of graphs at this stage. Then, a set of restrictions is applied to all possible graph edges for creating a graph with as less fake edges as possible, while preserving as many true edges as possible.

\begin{figure}[!ht]
    \centering
    \includegraphics[width=\linewidth]{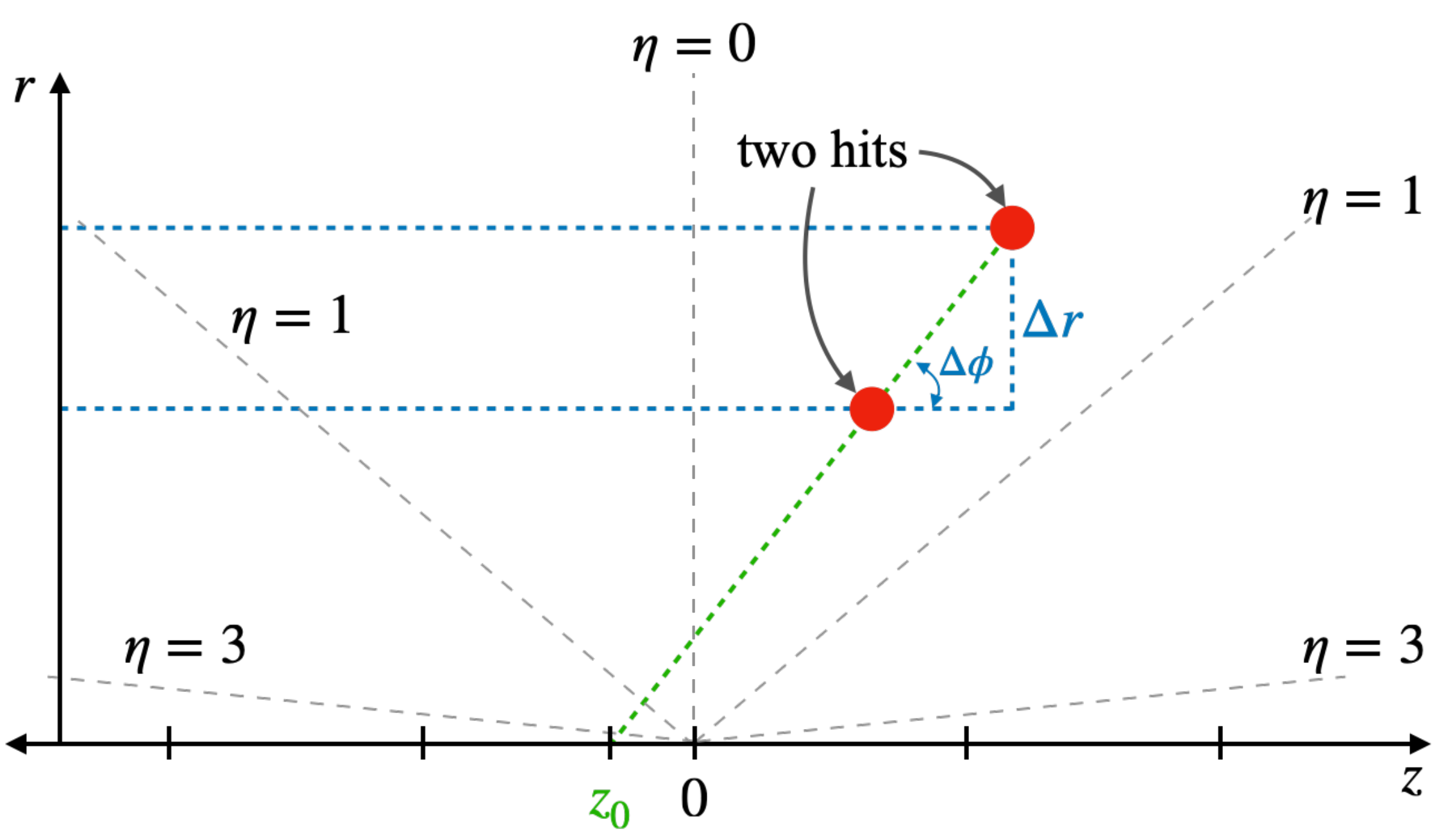}
    \caption{A sketch of the cylindrical coordinate system for particle collisions. The beam is along the $z$-axis and the particles collide near $z$=0. The $r$ axis is the projection of the transverse ($x$-$y$) plane.}
    \label{fig:coordinates}
\end{figure}

These restrictions are defined using a cylindrical coordinate system, which is widely used in High Energy Physics to leverage the symmetries of the detectors. We follow the same convention and present some of the definitions visually in Fig.~\ref{fig:coordinates} for further clarification.

As the first step of the hit graph construction, only the edges that connect nodes of consecutive detector layers are considered. Then, edges with pseudorapidity ($\eta = -\ln(\tan(\phi / 2))$ larger than 5 are eliminated. Pseudorapidity is a measure of angle to the z-axis used in High Energy Physics.

Next, the ratio of difference in $\phi$ to $r$ ($\Delta \phi/\Delta r$), where $\phi$ is the angle to the z-axis, is required to be smaller than $6\times10^{-4}$. Finally, a z intercept ($z_0$) of all edges is required to be smaller than 200 mm to eliminate highly oblique edges. 

Pseudo-code of the algorithm is also presented in Algorithm~\ref{alg:prepare}. Detailed plots of these selections are presented in Appendix~\ref{appendix:preprocessing}.

\begin{algorithm}
\caption{The pre-processing of an event}\label{alg:prepare}
\begin{algorithmic}
\Procedure{prepare}{}
\For{\texttt{all hits}}
\If{hit in Barrel Region}
    \State keep hit
\EndIf
\EndFor
\For{\texttt{all particles}}
\If{$pT > 1$ GeV}
    \For{\texttt{each hit of particle}}
        \State \texttt{\texttt{X[hit\_id]} = [$r$, $\phi$, $z$]}
    \EndFor
\EndIf
\EndFor
\State \textbf{set} \texttt{Ri, Ro and y = 0}
\For{\texttt{hit pairs of consecutive detector layers}}
    \If{$\eta < 5$}
    \If{$\Delta \phi/\Delta r < 0.0006$ and $z_0 < 100$ mm}
        \State \texttt{Ri[hit\_0\_id, segment\_id] = 1}
        \State \texttt{Ro[hit\_1\_id, segment\_id] = 1}
        \If{particle\_0\_id == particle\_1\_id}
            \State \texttt{y[segment\_id] = 1} 
        \EndIf
    \EndIf
\EndIf
\EndFor
\State \textbf{return} \texttt{X, Ri, Ro, y}
\EndProcedure
\end{algorithmic}
\end{algorithm}

In total, 100 graphs from 100 events are obtained with this method. The graph production is done with a 99\% efficiency and a purity of 51\%, which are defined as: 

\begin{equation}
    \mbox{Efficiency} = \frac{\mbox{\# of selected true track segments}}{\mbox{\# of initial track segments}},
\end{equation}

\begin{equation}
    \mbox{Purity} = \frac{\mbox{\# of selected true track segments}}{\mbox{\# of selected track segments}}.
\end{equation}

After the graph construction, the dataset is stored in form of 4 matrices; $X \in {\rm I\!R}^{N_V \times 3}$ stores 3 spatial coordinates of all nodes (in cylindrical coordinates; $r,\phi,z$), $R_i$ and $R_o$ ($R_i, R_o \in \{0,1\}^{N_V \times N_E}$) store input and output nodes of all edges, and $y \in \{0,1\}^{N_E}$ stores the labels of edges. Their definitions can be seen below.

\begin{equation}
        R_{i}^{\,jk} = 
\begin{cases}
    1, & \text{\, if $k^{th}$ edge is input of $j^{th}$ node}  \\
    0, & \text{\, otherwise}
\end{cases}
\end{equation}

\begin{equation}
        R_{o}^{\,jk} = 
\begin{cases}
    1, & \text{\, if $k^{th}$ edge is output of $j^{th}$ node}  \\
    0, & \text{\, otherwise}
\end{cases}
\end{equation}

\begin{equation}
    \, y_{k} \,\,\, = 
\begin{cases}
    1, & \begin{aligned}
    \text{if nodes of $k^{th}$ edge belong to}\\
    \text{same particle}
    \end{aligned} \\
    0, & \text{otherwise}
\end{cases}
\end{equation}

\begin{figure}[!b]
    \centering
    \includegraphics[width=\linewidth]{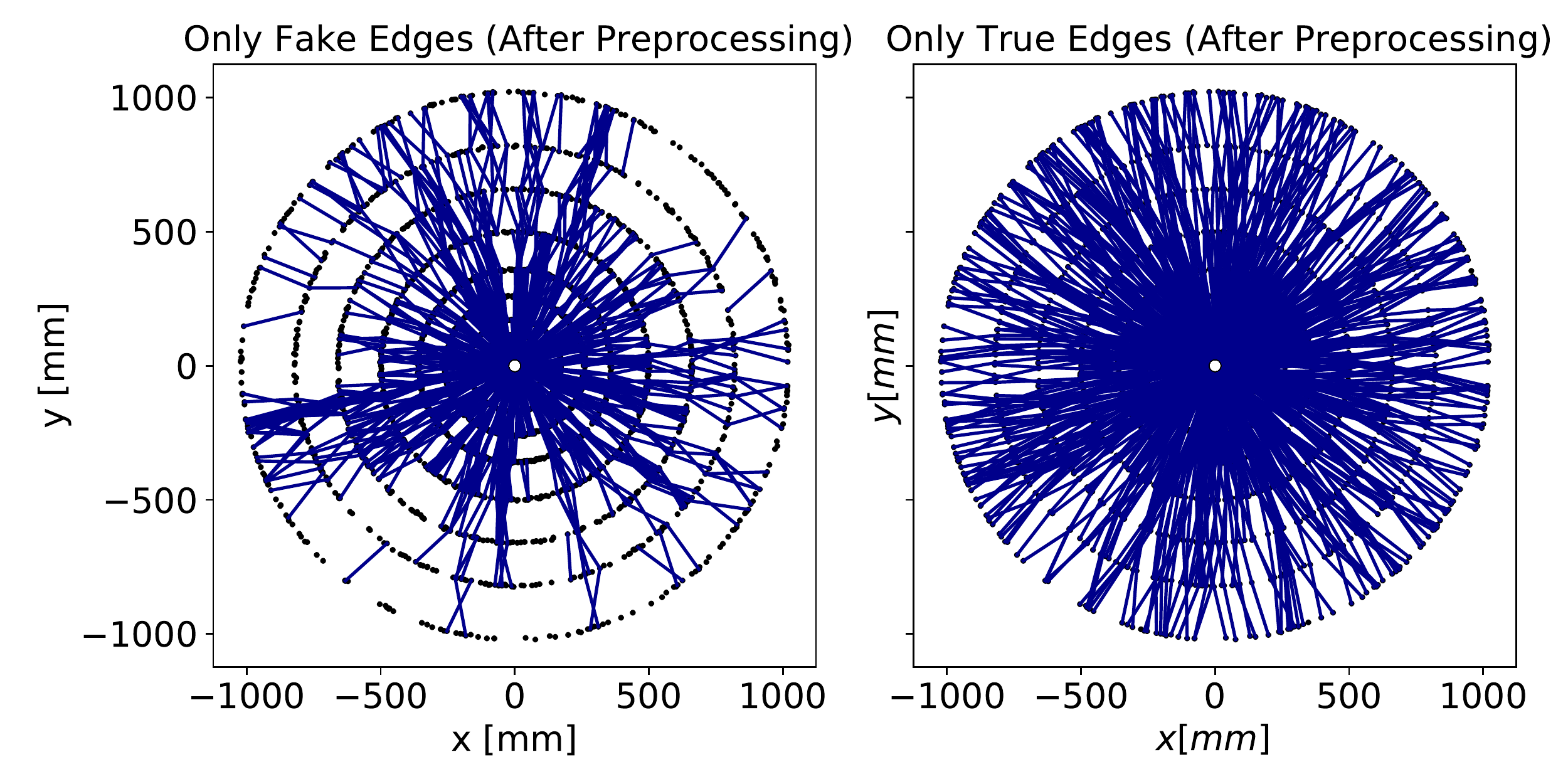}
    \caption{Graphs are produced with the pre-processing of each event. 2D projection of hits, fake and true edges of an event are plotted. All hits are plotted with black circles. Fake (on the left) and true (on the right) edges of a graph are plotted in the Cartesian coordinates (transverse plane). There are 5162 true and 5508 fake edges of this event.}
    \label{fig:graph}
\end{figure}

Constructed graphs have 8784 $\pm$ 1877 edges ($N_E$) and 5583 $\pm$ 804 nodes ($N_V$) on average. An example graph showing the fake and true edges is presented in Fig.~\ref{fig:graph}.

The pre-proprecessing method is identical to the one used in HEP.TrkX project (\cite{HepTrkX}), except the $pT$ restriction which is used to reduce total number of particles in an event. This is done intentionally to compare our results with the classical equivalent model.

\section{The Hybrid Quantum-Classical Graph Neural Network Model}
\label{hgnn}

A Graph Neural Network (GNN) is a Neural Network model that acts on features of the graph, such as nodes, edges or global features (\cite{velickovic2018graph}). GNNs have shown great success in many occasions for node and graph classification and link prediction (\cite{wu_comprehensive_2021}). Their success led to applications in High Energy Physics for many problems such as track and particle flow construction (\cite{HepTrkX, ju_graph_2020, shlomi_graph_2020, biscarat_towards_2021, pata_mlpf_2021}). This situation attracted interest from the Quantum Machine Learning community to develop Quantum Graph Neural Networks for different applications (\cite{verdon_quantum_2019, chen_hybrid_2021}). 

The Hybrid Quantum Classical Graph Neural Network (QGNN) model that we propose takes a graph as the input and gives a probability as the output for all edges of the initial graph. The QGNN builds up an Attention Passing Graph Neural Network model proposed by \cite{velickovic2018graph}, following the same strategy as the HEP.TrkX project of \cite{HepTrkX}. In contrast to the classical GNN approach, we add a Quantum Neural Network (QNN) layer to Multi Layer Perceptrons (MLP). 

The QGNN consists of 3 parts. The first one is the Input Network, whose task is to increase the dimension of the input data. It takes the spatial coordinate information (e.g. 3 cylindrical coordinates) and passes them through a single fully connected Neural Network layer with sigmoid activation and an output size corresponding to the hidden dimension size ($N_D$). Then, these new data points are concatenated ($\oplus$) to form the initial node feature vector, where $v \in {\rm I\!R}^{N_V \times (3+N_D)}$.

\begin{equation}
  v = x \oplus \text{\Large$\phi$}_{FC}(x)
    \label{eq:input_network}
\end{equation}

\begin{figure}[!ht]
    \centering
    \includegraphics[width=\linewidth]{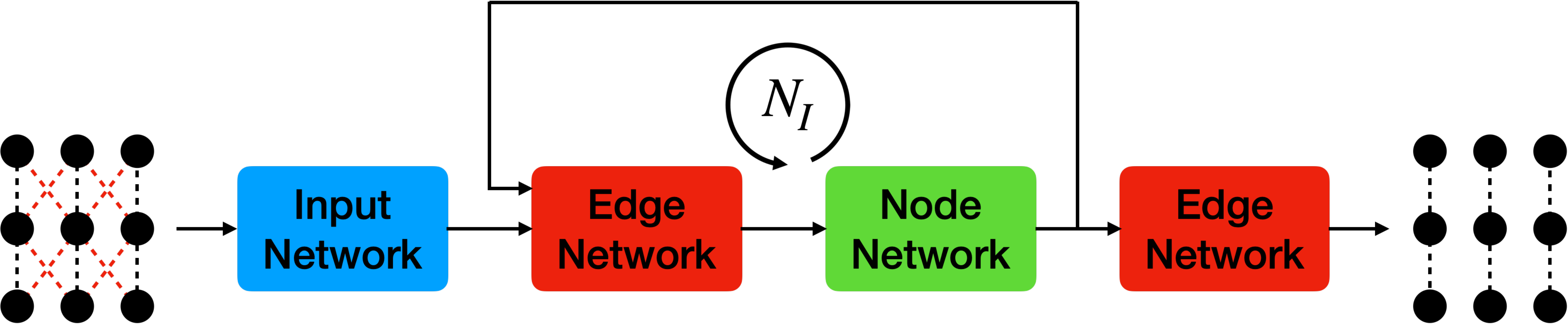}
    \caption{Schematic of the QGNN architecture. The pre-processed graph is fed to an Input Network, which increases the dimension of the node features. Then, the graph's features are updated with the Edge and Node Networks iteratively, number of iterations ($N_I$) times. Finally, the same Edge Network is used one more time to extract the edge features of the graph that predicts the track segments. There is only one Edge Network in the pipeline, two Edge Networks are drawn only for visual purposes. The pipeline is adapted from~\cite{HepTrkX}.}
    \label{fig:hgnn}
\end{figure}

Next, the node feature vector is fed to Edge and Node Networks, which process the graph iteratively in order to obtain a final edge probability value ($e$) for each of the edges. During this process, the same Edge and Node Network is sequentially executed for a predetermined number of iterations ($N_{I}$) times and finally the same Edge Network is used one more time to obtain final edge probabilities ($e \in [0,1]^{N_E} $). This pipeline is summarized with a simple drawing in Fig.~\ref{fig:hgnn}.

\subsection{The Edge Network}
\label{en}
The Edge Network takes pairs of nodes into account and returns the probability for those two nodes to be connected. Initially, the connectivity of each pair of nodes is given by the connectivity matrices $R_i$ and $R_o$. Using these matrices, node feature vectors $b_o$ and $b_i$ of all initially connected edges, or so called \textit{doublets} ($b_o \oplus b_i$), are obtained. 

\begin{align}
    b_o^{\,k} = \sum_{j=1}^{N_V} R_{o}^{\,jk}v_j &&
    b_i^{\,k} = \sum_{j=1}^{N_V} R_{i}^{\,jk}v_j
\end{align}

The feature vectors of input and output nodes of each edge are concatenated to be fed into a Hybrid Neural Network (HNN, $\text{\Large$\phi$}_{EdgeNetwork}$). The HNN returns edges features $(e)$, which are the probabilities for each edge, to be part of a real trajectory or not. Next, the edge features are passed to the Node Network.

\begin{equation}
    e_k = \text{\Large$\phi$}_{EdgeNetwork}(b_o^{\,k} \oplus b_i^{\,k})
    \label{eq:edge_network}
\end{equation}

\subsection{The Node Network}
\label{nn}
The Node Network builds up on the edge feature matrix given by its predecessor, the Edge Network. Based on this input information, the node features are updated. In this case, a combination of each node of interest and its neighbors from upper and lower detectors is created, forming a \textit{triplet}. Here, the node features of the neighbors' are scaled with the corresponding edge features.

\begin{align}
   v^\prime_{j,input} = \sum_{k=1}^{N_E} e_{k}R_{i}^{\,jk}b_{o}^k &&
   v^\prime_{j,output} = \sum_{k=1}^{N_E} e_{k}R_{o}^{\,jk}b_{i}^k
\end{align}

Similar to the Edge Network, the triplet is fed to a Hybrid Neural Network ($\text{\Large$\phi$}_{NodeNetwork}$).

\begin{equation}
    v_j \coloneqq \text{\Large$\phi$}_{NodeNetwork}(v^{\prime}_{j,input} \oplus v^{\prime}_{j,output} \oplus v_j)
    \label{eq:node_network}
\end{equation}

This time, the HNN returns new node features $v$. The updated features are passed again to the Edge Network and this process is repeated for $N_I$ times. This allows the aggregation of information from farther nodes of the graphs and updates the hidden features accordingly. 

\begin{figure*}[!ht]
    \centering
    \includegraphics[width=\linewidth]{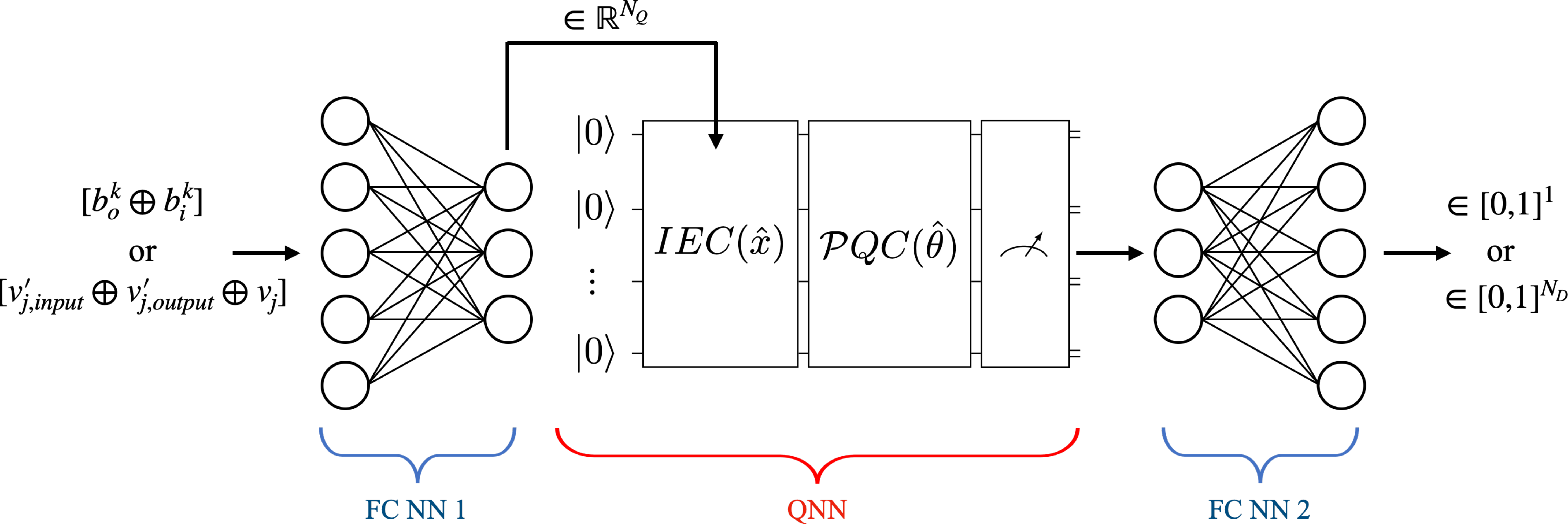}
    \caption{The Hybrid Neural Network (HNN) architecture. The input is first fed into a classical fully connected Neural Network (FC NN) layer with sigmoid activation. Then, its output is encoded in the QNN with the Information Encoding Circuit (IEC). Next, the Parametrized Quantum Circuit (PQC) applies transformations on the encoded states. The output of QNN is obtained as expectation values for each qubit that is measured. A final FC NN layer with sigmoid activation is used to combine the results of different qubit measurements. The same HNN architecture is used in Edge (upper input and output dimension) and Node Networks (lower input and output dimension) with different parameters. The input and output dimension sizes change according to the network type. Details of the dimensions of each layer are given in Table~\ref{tab:hnn}.}
    \label{fig:hnn}
\end{figure*}

\subsection{The Hybrid Neural Network}
\label{hnn}

Our approach employs Hybrid Neural Networks (HNNs), which combine both classical and quantum layers. The HNN starts with a single fully connected neural network (FC NN 1) layer with sigmoid activation. The output dimension of this layer is equal to number of qubits ($N_Q$) used by the quantum layer. Then, the output of the FC NN 1 is used in the encoding step of the QNN. Finally, the measurements of the QNN is fed to another FC NN with sigmoid activation, which has the output dimension of 1 (in the case of Edge Network) or hidden dimension size ($N_D$) (in the case of Node Network). This architecture, as presented in Fig.~\ref{fig:hnn}, allows full flexibility in hidden dimension size, number of qubits and the type of the QNN. Details of input and output dimensions of all layers can be seen in Table~\ref{tab:hnn}.

\begin{table}[!ht]
\caption{Input and Output dimensions of layers used in the HNN. QNN has the output dimension of 1 if the circuit measures only one qubit (e.g. MPS and TNN). QNN has the output dimension of $N_Q$ if all qubits are measured (e.g. Circuit 10 and Circuit 19).}
\label{tab:hnn}       
\begin{tabular}{lccc}
\hline\noalign{\smallskip}
Layer & I/O & Edge Network  & Node Network  \\
\noalign{\smallskip}\hline\noalign{\smallskip}
FC NN 1 & \begin{tabular}{@{}c@{}}
                   Input\\
                   Output\\
                 \end{tabular}
    & \begin{tabular}{@{}c@{}}
                   $2\times(3+N_{D})$\\
                   $N_{Q}$\\
                 \end{tabular}
    & \begin{tabular}{@{}c@{}}
                   $3\times(3+N_{D})$\\
                   $N_{Q}$\\
                 \end{tabular}\\
\noalign{\smallskip}\hline
QNN & \begin{tabular}{@{}c@{}}
                   Input\\
                   Output\\
                 \end{tabular}
    & \begin{tabular}{@{}c@{}}
                   $N_{Q}$\\
                   1 or $N_{Q}$\\
                 \end{tabular}
    & \begin{tabular}{@{}c@{}}
                   $N_{Q}$\\
                   $N_{Q}$\\
                 \end{tabular}\\
\noalign{\smallskip}\hline
FC NN 2& \begin{tabular}{@{}c@{}}
                   Input\\
                   Output\\
                 \end{tabular}
    & \begin{tabular}{@{}c@{}}
                   1 or $N_{Q}$\\
                   1\\
                 \end{tabular}
    & \begin{tabular}{@{}c@{}}
                   $N_{Q}$\\
                   $N_{D}$\\
                 \end{tabular}\\
\noalign{\smallskip}\hline
\end{tabular}
\end{table}

The QGNN model is experimented with different quantum layers to understand potential benefits. These type of quantum models with parametrized quantum circuits have been called differently in the literature (\cite{barren, farhi_classification_2018, benedetti_parameterized_2019, mitarai_quantum_2018, mcclean_theory_2016, romero_variational_2021}). Here, we use the name Quantum Neural Network (QNN) as we use them in a similar fashion to Neural Network layers. 

The QNN of our choice consists of three consecutive parts. An information encoding circuit (IEC) encodes classical data to states of the qubits followed by a parametrized quantum circuit (PQC) that is applied to transform these states to their optimal location on the Hilbert Space. Finally, measurements are performed along the z-axis with the $\sigma_z$ operator.

Information encoding has a significant effect on the training capacity of QNN models~\cite{schuld_effect_2021}, therefore a lot of attention is required when deciding on how to do it. We employ angle encoding, because it provides an encoding which uses significantly less gates compared to others, e.g. amplitude encoding, and it needs almost no classical processing (\cite{Larose2020}). 

\begin{figure}[!ht]
    \centering
    \includegraphics{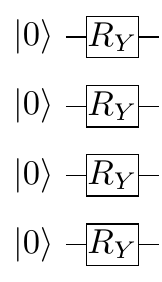}
    \caption{Angle encoding quantum circuit of a four dimensional feature vector with respect to the $y$-axis.}
    \label{fig:angle_encoding}
\end{figure}

Encodings such as amplitude encoding allow encoding of classical information by using significantly less qubits, but this advantage is usually reverted by the number of gates required to build the circuit. For example, the amplitude encoding of a feature vector $x \in {\rm I\!R}^{n}$ only uses $log_2\,n$ qubits but needs $4^n$ single and two qubit gates. On the other hand, the angle encoding requires $n$ number of qubits and a single qubit gate per qubit~(\cite{leymann_bitter_2020}). This allows an easier implementation and experimentation with angle encoding. An angle encoding of a four dimensional feature vector can be performed, e.g., with the circuit given in Fig.~\ref{fig:angle_encoding}.

The QNN encodes classical incoming information on the qubits via rotational gates in the desired axis using angle encoding. In order to obtain a unique and bijective representation of the classical data, the rotation angle is mapped between $\theta \in [0,\pi]$ due to the periodicity of the cosine function. This is relevant since the expectation value is taken with respect to the $\sigma_z$-operator at the end of the circuit execution.

The PQC is the part of the QNN model that is going to be tuned in order to provide the desired output. As in classical Neural Network layers, those initially randomly assigned variables are optimized during training to fit the certain training objective, i.e., minimize the overall loss function. In order to achieve a good training performance, choosing a good combination of IEC and PQC is essential. Although there are many practical and theoretical work to understand this better (\cite{qc-assesment, Leyton-Ortega2021, Hubregtsen2021}), our current understanding of which combination works for which task is still limited (\cite{schuld_effect_2021}). We therefore try to cover a range of PQCs and fix the IEC to a specific angle encoding to provide more controlled results. We consider two types of PQCs.

\begin{figure*}[!ht]
\centering
\subfigure[Matrix Product State (MPS) circuit. Adapted from~\cite{mps}.]{\includegraphics[height=0.225\linewidth]{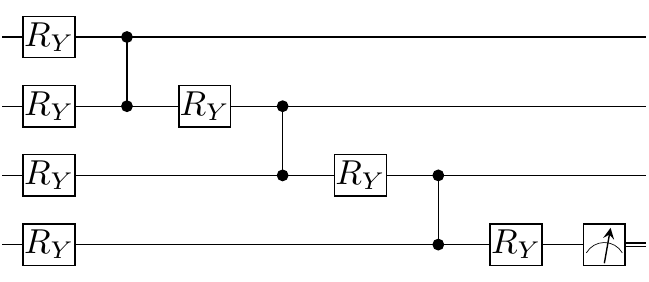}}\label{fig:MPS}
\hfill
\subfigure[Tree Tensor Network (TTN) circuit. Adapted from~\cite{ttn}.]{\includegraphics[height=0.225\linewidth]{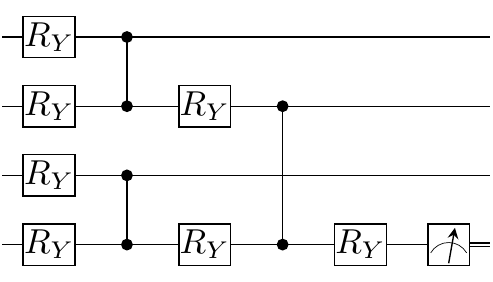}}\label{fig:TTN}
\subfigure[Circuit 19 in four qubits and single layer configuration. Adapted from~\cite{qc-assesment}.]{\includegraphics[height=0.225\linewidth]{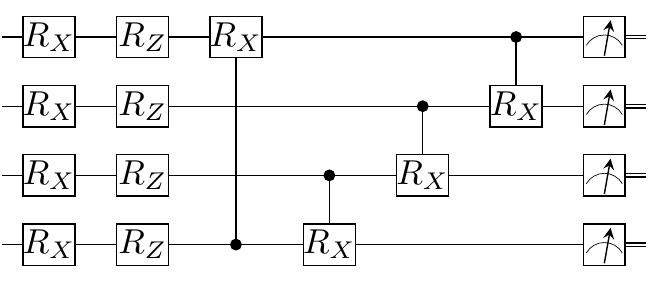}}\label{fig:c19}
\hfill
\subfigure[Circuit 10 in four qubits and single layer configuration. Adapted from~\cite{qc-assesment}.]{\includegraphics[height=0.225\linewidth]{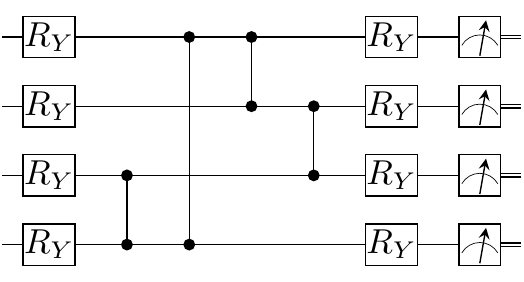}}\label{fig:c10}
\caption{Hierarchical (a, b) and layered (c, d) PQCs used in the HNN. (a) MPS applies two-qubit gates with a ladder-like architecture, (b) while TTN uses a tree-like architecture implement $R_Y$ and $CZ$ gates. MPS and TTN circuits measure only one qubit at the end. (c) Circuit 19 employs $R_X$, $R_Z$ and $CRX$ gates in a nearest neighbour fashion, (d) while Circuit 10 do this with $R_Y$ and $CZ$ gates. Circuit 10 and Circuit 19 can be extended to any number of layers by repating the circuit. Circuit 10 and Circuit 19 measure all qubits available.} \label{fig:pqcs}
\end{figure*}

The first PQC type consists of circuits with a hierarchical architecture. Matrix Product State (MPS) (\cite{mps}) and Tree Tensor Network (TTN) (\cite{ttn}) inspired circuits belong to this group. However, these PQCs measure only one qubit. Thus, they are only implemented in case of the Edge Network as a multi-dimensional output is needed for the Node Network. Examples of MPS and TTN circuits can be seen in Fig.~\ref{fig:pqcs}.

The second type of PQCs are more common in the QML literature. They consist of layers of parametrized gates that are generally followed by controlled operations. These circuits act on all qubits fairly, meaning all qubits can be measured to obtain information. This makes them suitable for both Edge and Node Networks. Another important advantage of this type of PQCs is having different descriptors in the literature, which allows to compare different properties such as expressibility and entanglement capacity. In this work, two circuits with different expressibility and entanglement capacity were chosen from~\cite{qc-assesment}, namely Circuit 10~(Fig.~\ref{fig:c10}) and Circuit 19~(Fig.~\ref{fig:c19}).

This work compares the two different configurations of PQCs. Models with the labels \texttt{circuit 10} and \texttt{circuit 19} use the same circuits with different initial parameters for the Edge and Node Networks, as it was done in previous comparisons. While the models with \texttt{TTN-10} and \texttt{MPS-10} labels use \texttt{circuit 10} for the Node Network, a TTN or an MPS type of PQC for the Edge Network. These definitions are also presented in Table~\ref{tab:1}.

\begin{table}[!ht]
\centering
\caption{Labels of PQC settings used in the HNN.}
\label{tab:1}       
\begin{tabular}{lll}
\hline\noalign{\smallskip}
Label &  Edge Network & Node Network  \\
\noalign{\smallskip}\hline\noalign{\smallskip}
\texttt{circuit 10} & Circuit 10 & Circuit 10 \\
\texttt{circuit 19} & Circuit 19 & Circuit 19 \\
\texttt{MPS-10} & MPS & Circuit 10 \\
\texttt{TTN-10} & TTN & Circuit 10 \\
\noalign{\smallskip}\hline
\end{tabular}
\end{table}

Expressibility measures a PQC's ability to explore the Hilbert Space~\cite{qc-assesment}. It is a numerical method that samples two random states from a given PQC. Fidelities of these states are computed and a distribution ($\hat{P}_{PQC}(F ; \theta)$) is obtained after collecting many samples (e.g. 5000 samples for four qubits). Then, this process is repeated by sampling Haar random states. Finally, two distributions are compared using Kullback Leibler divergence ($D_{KL}$). Expressibility ($E$) is expressed as;

\begin{equation}
    {\rm E} = D_{KL} (\hat{P}_{PQC}(F ; \theta) \parallel P_{Haar}(F))
\end{equation}

The value of Expressibility is less for more expressive circuits. In order to avoid confusion $E' = -log_{10}(E)$ will be used as Expressibility so that, the Expressibility value (E') increases with more expressive circuits~\cite{Hubregtsen2021}.

Similarly, Entanglement Capability is a numerical method that quantifies a PQCs ability to produce entangled states~\cite{qc-assesment}. It averages the Meyer-Wallach entanglement measure ($Q$) over many random samples obtained from the PQC (e.g. 5000 samples for four qubits). For example, a fully entangled two qubit state $\left( \ket{\Psi} = \frac{\ket{00}+\ket{11}}{\sqrt{2}} \right)$ has $Q(\ket{\Psi}) = 1$ and a state with no entanglement (e.g. $\ket{\Phi} = \ket{01}$) has $Q(\ket{\Phi})=0$. Entanglement Capability (Ent) is expressed as;

\begin{equation}
    Ent = \frac{1}{\norm{S}} \sum_{\theta_i \in S} Q(\ket{\psi_{\theta_i}})
    \label{eq:ent}
\end{equation}

Expressibility and Entanglement Capability are descriptors that allow comparison of layered PQCs. Expressibility and Entanglement Capability improve with more layers and will be a point of interest in our discussions. For example, Circuit 10 has less Expressibility and Entanglement Capability compared to Circuit 19, with the same number of qubits and layers. This is because Circuit 10 has only $R_Y$ and $CZ$ gates compared to additional $R_Z$ and $CR_X$ gates of Circuit 19, which brings additional degrees of freedom. The differences between these circuits will be another point of interest in our comparisons to better understand the behaviour of PQCs. Expressibility and Entanglement Capability of Circuit 10 and Circuit 19 is presented in Appendix~\ref{appendix:exp-ent}.

The reason behind analyzing PQCs under two types is related to their response to scaling with increasing number of qubits and layers. We use gradient-based optimizers due to the hybrid nature of QGNN model. Gradient-based optimizers require the model to produce strong enough gradient signals to be able to explore the loss landscape. This might become a problem when a model is scaled. Barren Plateaus are the name given to flattening of the loss landscape~(\cite{barren}). They appear in models where gradients vanish exponentially with an increasing model size and they are one of the greatest challenges in training Variational Quantum Algorithms (VQAs)~(\cite{cerezo_variational_2021}). In general, layered-type PQCs suffer from Barren Plateaus, which makes them hard or even impossible to train for a large amount of qubits and layers. On the other hand, the absence of Barren Plateaus were shown for some PQCs with hierarchical architectures, such as Quantum Convolutional Neural Networks (QCNNs)~(\cite{pesah2020absence}) and TTNs~(\cite{zhao_analyzing_2021}). Because of this, comparing these two types of PQCs have great importance to better understand the behaviour of hybrid models at large scales.

\subsection{Training the Network}
\label{training}

Training hybrid quantum-classical neural networks requires software that can differentiate both types of networks. Pennylane by \cite{pennylane} is one of the most popular open-source tools that provides this feature. Pennylane was used along with PyTorch (\cite{pytorch}) during the early stages of this work. However, this combination turned out to be too slow to handle both the dataset and the model. A computational speed-up in training has been achieved using Qulacs (\cite{suzuki2020qulacs}). Although it provided faster training, it was still not enough. Finally, the combination of  Cirq, Tensorflow and Tensorflow Quantum (\cite{cirq, abadi2016tensorflow, tfq}) produced the optimal scenario, in which we were able to reduce the training times to less than a week. The quantum circuit simulations are performed with only taking analytical results into account, i.e. without sampling the quantum circuits. Although analytical results do not reflect hardware conditions, we made this choice in order to obtain results in a reasonable amount of time.

The 100 events selected from the dataset are separated randomly with a 50/50 ratio to be used as training and validation sets. Models are trained using the binary cross entropy loss function, given in Eq.~\ref{eq:loss}, where $y_i$ is the truth label and $\hat{y}_i$ is the model prediction for an edge. 

\begin{equation}
    L = -\frac{1}{N_E} \sum_{i=1}^{N_E} y_i \text{log}\,(1-\hat{y}_i) + (1-y_i)\text{log}\,{\hat{y}_i}
    \label{eq:loss}
\end{equation}

The Adam optimizer (\cite{kingma_adam_2017}) with a learning rate of 0.01 is used to train all trainable parameters of the hybrid model. The learning is done with a batch size of 1 per graph and continued up-to 10 or 20 epochs depending on model complexity. All models are trained for 5 independent initializations and their mean is presented in all results.

\begin{figure*}[!ht]
\centering
\subfigure[Axis of angle embedding comparison.]{\includegraphics[width=0.48\textwidth]{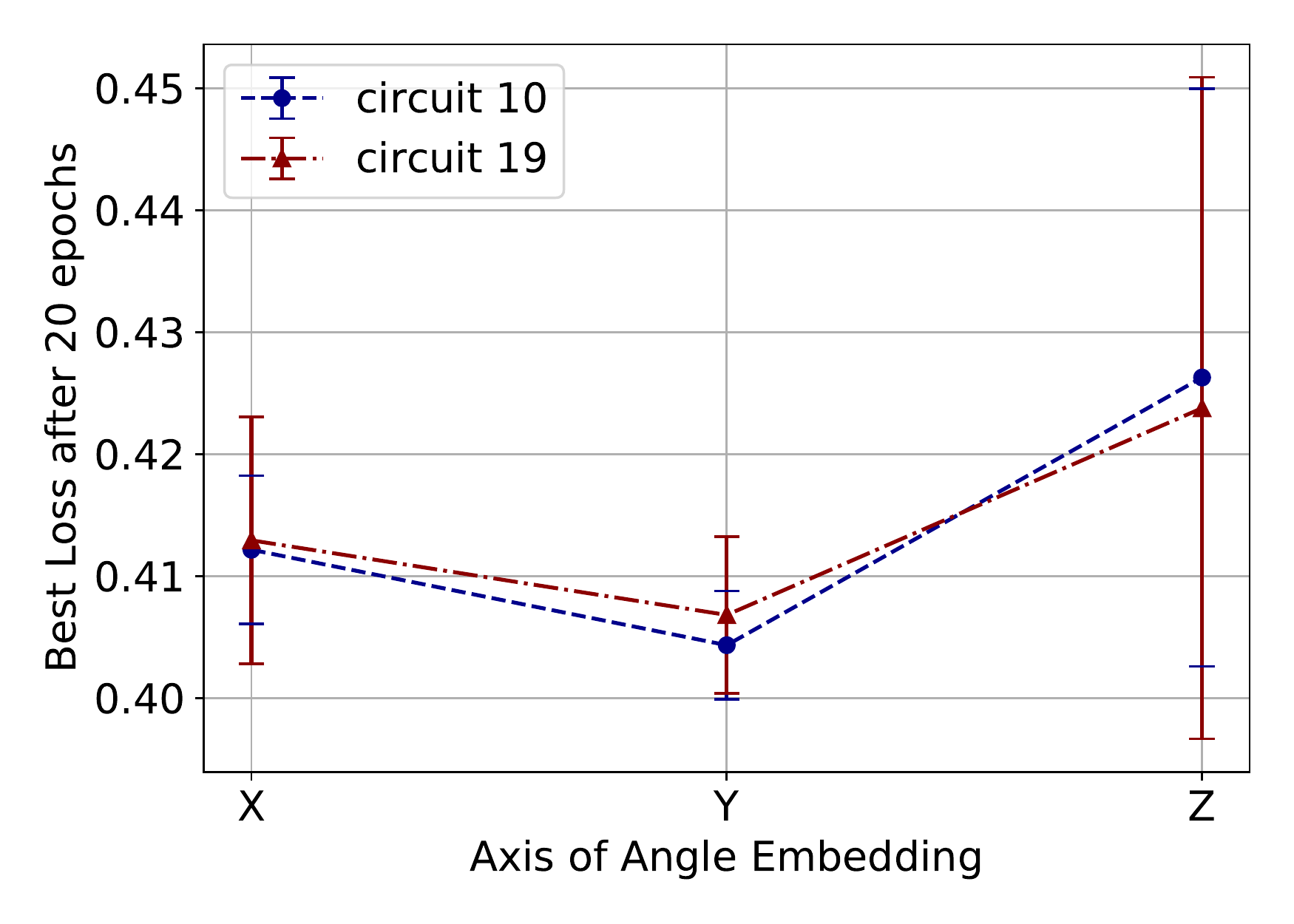}}\label{fig:embedding_comparison}
\hfill
\subfigure[Number of layers ($N_L$) comparison.]{\includegraphics[width=0.48\textwidth]{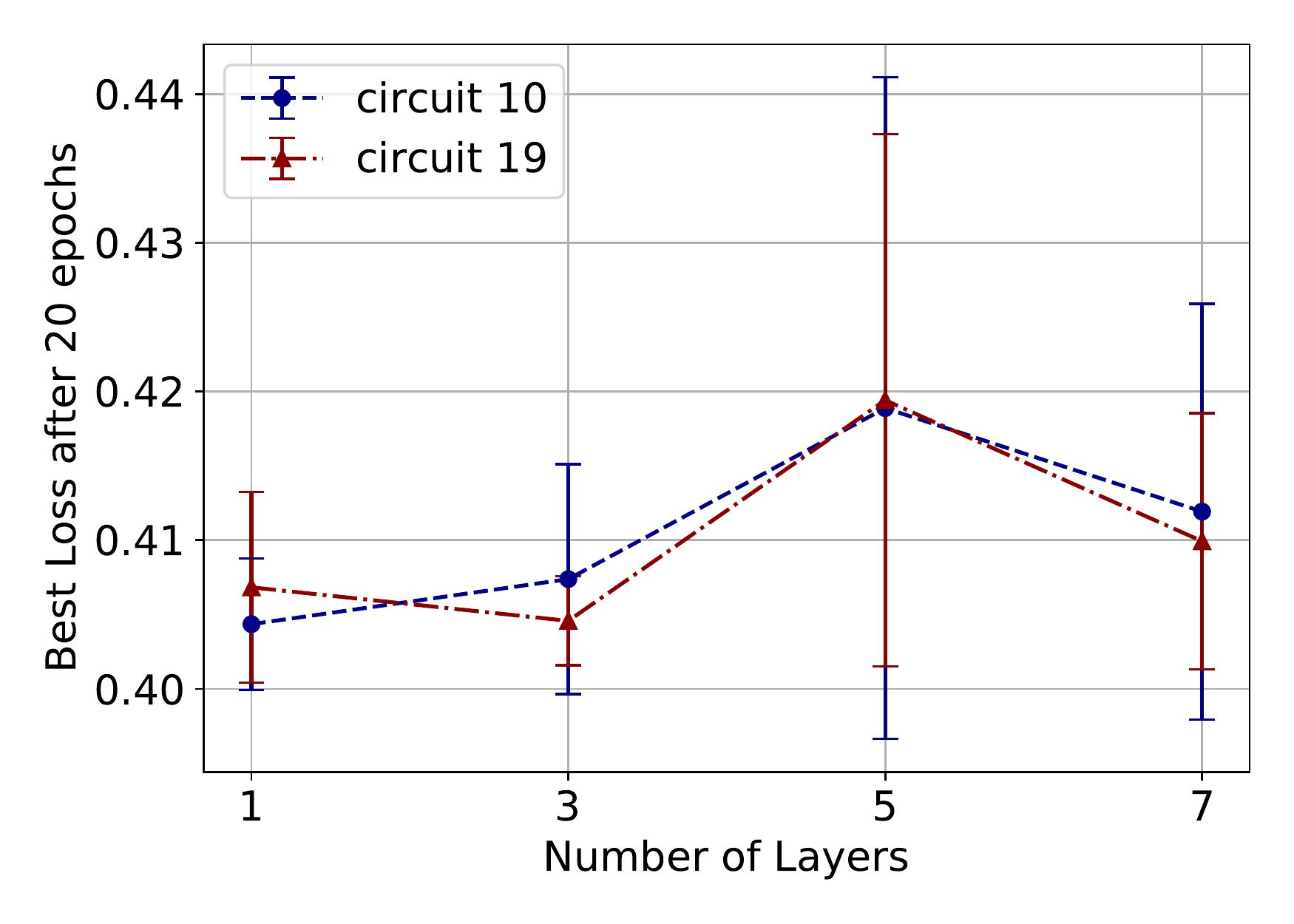}}\label{fig:layer_comparison}
\subfigure[Number of iterations ($N_{I}$) comparison.]{\includegraphics[width=0.48\textwidth]{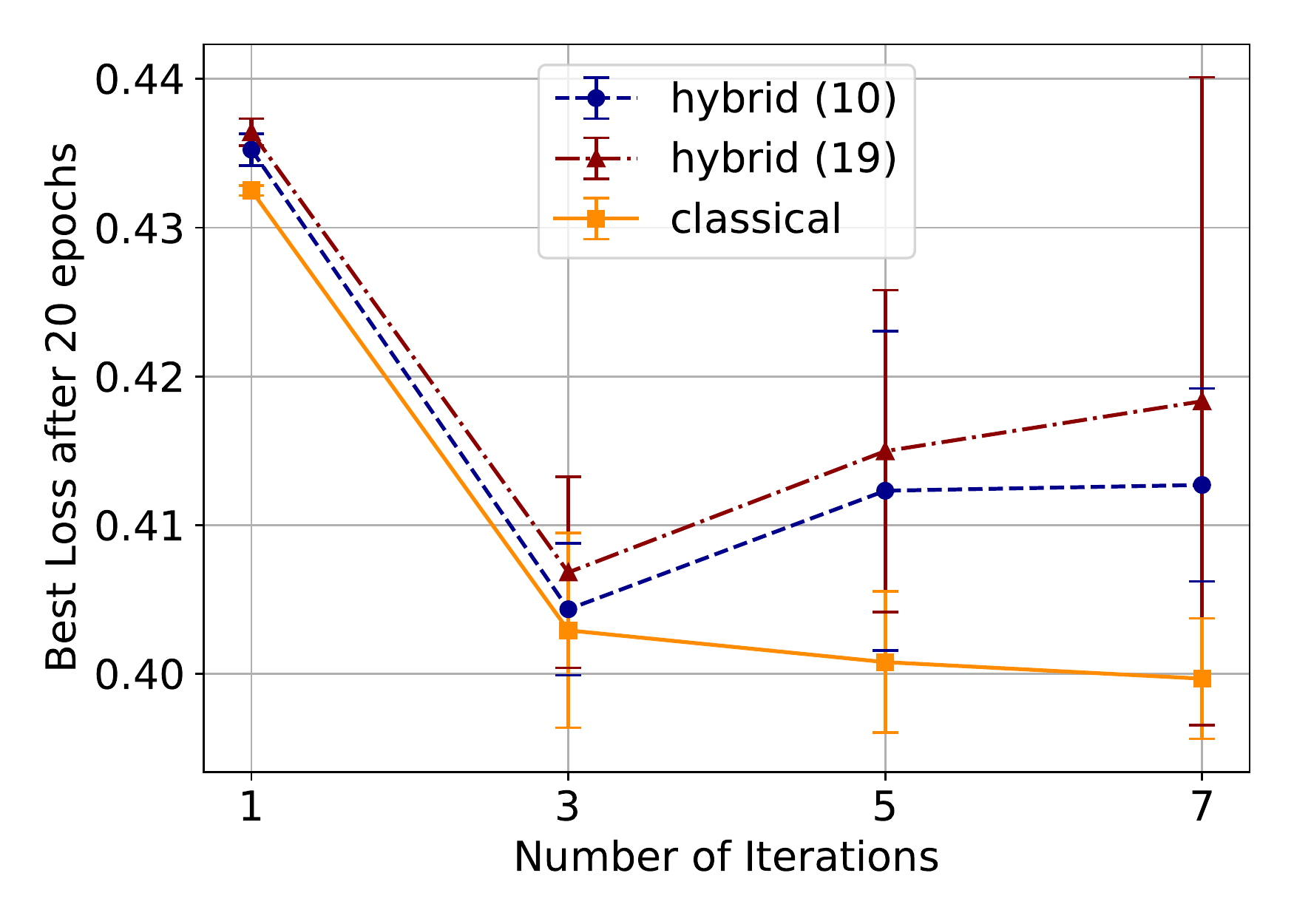}}\label{fig:iteration_comparison}
\hfill
\subfigure[Hidden dimension size ($N_D=N_Q$) comparison.]{\includegraphics[width=0.48\textwidth]{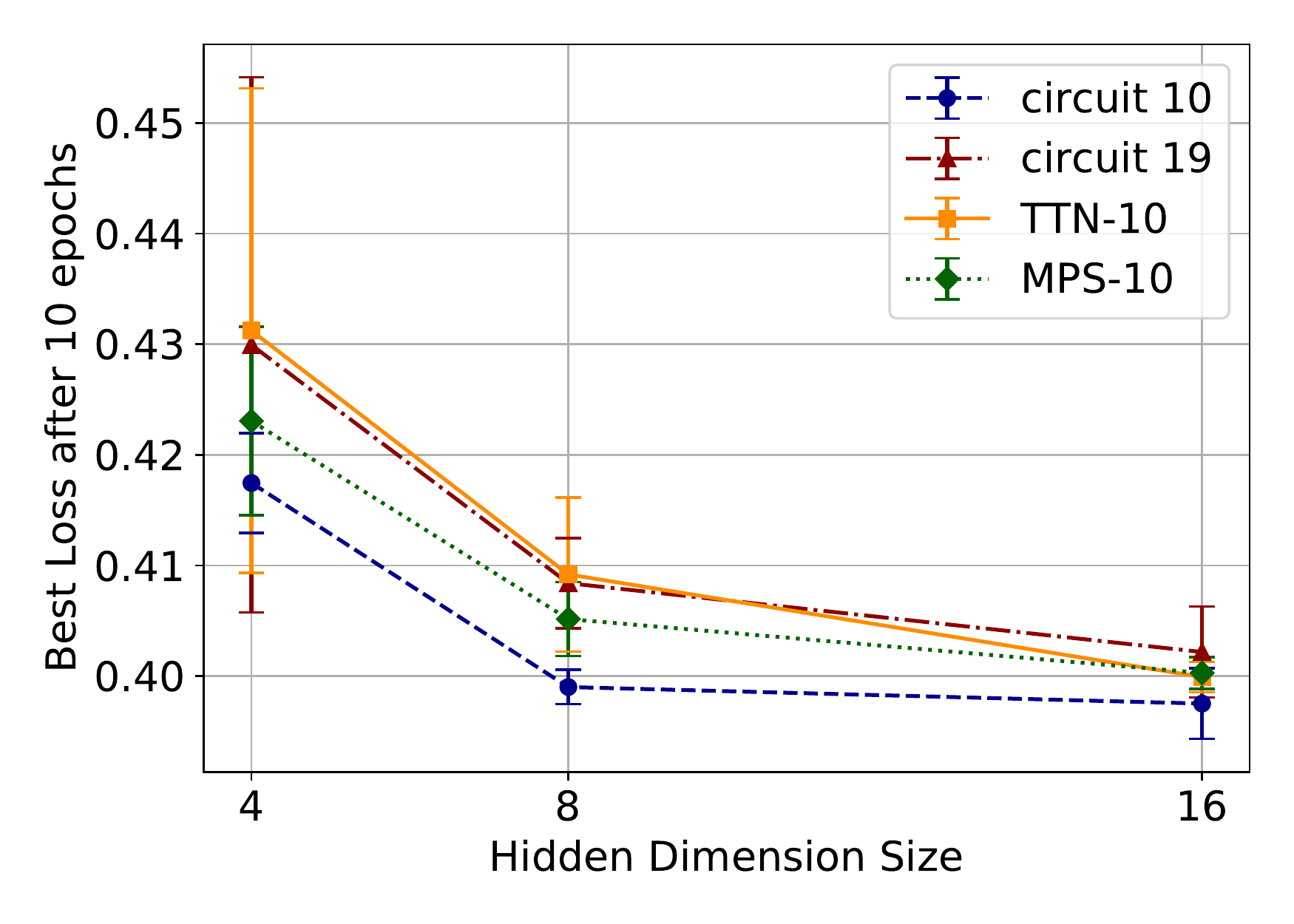}}\label{fig:dimension_comparison}
\caption{Best validation loss comparison with respect to different parameters of the hybrid GNN model.  (a) The axis of angle embedding comparison considers the best loss obtained for different embedding axes by setting $N_D=N_Q=4$, $N_L=1$ and $N_I=3$. (b) The number of layers comparison considers the best loss for various numbers of layers ($N_L$) by setting $N_D=N_Q=4$ and $N_I=3$. (c) The number of iterations comparison considers the best loss for different numbers of iterations ($N_{I}$) by setting $N_D=N_Q=4$ and $N_L=1$. (d) The hidden dimension size comparison considers the loss for different hidden dimension sizes ($N_D$) by setting $N_Q=N_D$, $N_L=1$ and $N_I=3$. 5 instances of all models with different initial parameters are trained for 10 or 20 epochs depending on complexity for each setting, and the mean of best losses are presented. The error bars represent the $\pm$ standard deviation of the best losses of all 5 runs.} \label{fig:comparisons}
\end{figure*}

\section{Results \& Discussion}
\label{results}

We trained the hybrid model with many configurations to explore the potential of the method. Here, we present four key comparisons of features that have a significant effect on the performance of the model.

First, the effect of angle embedding axis choice on the training performance of \texttt{circuit 10} and \texttt{19} is compared. Circuit 10 is a PQC that consists of $R_Y$ and $CZ$ gates, while circuit 19 is a PQC with $R_Z$, $R_X$ and $CRX$ gates. The comparison is made by setting number of qubits and number of hidden dimension size to 4 ($N_Q=N_D=4$). Then, number of layers is also set to 1 ($N_L=1$) and number of iterations is set to 3 ($N_I=3$). The best loss values of each model is plotted in Fig.~\ref{fig:embedding_comparison}. In both cases, the $x$ and $y$-axis embedding resulted in better loss values, compared to $z$-axis. The $z$-axis embedding requires deeper circuits to match other axes' representation capacity, since the measurements are taken with respect to the Pauli-Z operator. Because of this outcome, the $y$-axis angle embedding is considered for the rest of the results. Training curves of these comparisons are presented in Appendix~\ref{appendix:embedding}.

There are contrasting results in the literature on how expressibility and entanglement capacities affect the training performance. Recently, \cite{Hubregtsen2021} showed a positive correlation between expressibility and accuracy, while \cite{Leyton-Ortega2021} showed the opposite. They found that more expressive models perform worse and also overfit more. On the other hand, entanglement has been shown to limit the trainability of models depending on how it propagates in between qubits by \cite{Marrero2020} and \cite{zhang2020trainability}.  
To better understand the situation on our case, we tested two models with \texttt{circuit 19} and \texttt{circuit 10} with various number of layers. \texttt{circuit 19} has better expressibility and entanglement capacity compared to \texttt{circuit 10} (\cite{qc-assesment}). Best loss values obtained after a training with 20 epochs is plotted with respect to number of layers in Fig.~\ref{fig:layer_comparison} with $N_Q=N_D=3$ and $N_I=3$. We could not observe a significant difference between two models. This situation might be a result of using an encoder and decoder consisting of a fully connected neural network in the model, which could have compensated for the different expressive capacity of the models. However, increasing expressibility and entanglement capacity of both models resulted in worse performance in  both cases. In this way, our results are consistent with results of \cite{Leyton-Ortega2021}. This behaviour is thought to be the result of Barren Plateau formation (\cite{barren}). Training curves of these comparisons are presented in Appendix~\ref{appendix:layer}.

The number of iterations of a GNN is an important parameter that determines a model's performance (\cite{HepTrkX, velickovic2018graph}). It allows propagation of information to farther nodes. A comparison with $N_Q=N_D=3$ and $N_L=1$ is made with \texttt{circuit 10} and \texttt{circuit 19} and the results are presented in Fig.~\ref{fig:iteration_comparison}. Training results show that the best loss is obtained for $N_I=3$ for the hybrid cases. However, this is not the case in the classical case. \cite{ju_graph_2020} report an $N_I=8$ as the optimal value for their model with 128 hidden dimensions in their extended project Exa.TrkX. The increase in the value of the lowest loss with increasing number of iterations might be due to low expressive capacity of the whole model, as this comparison is made only with a $N_D=4$. Training curves of these comparisons are presented in Appendix~\ref{appendix:iter}.

In order to investigate how these hybrid models scale, their performances with respect to increasing the hidden dimension size and qubits are compared. This comparison is made with the choice of $N_{Q}=N_{D}$, $N_I=3$ and $N_L=1$. Two different configurations of PQCs are compared. Models with the labels \texttt{circuit 10} and \texttt{circuit 19} use the same circuits with different initial parameters for the Edge and Node Networks, as it was done in previous comparisons. While the models with \texttt{TTN-10} and \texttt{MPS-10} labels use Circuit 10 for the Node Network, a TTN or an MPS type of PQC for the Edge Network. These definitions are given in Table~\ref{tab:1}.

A comparison of the best losses is made after 10 epochs and presented in Fig.~\ref{fig:dimension_comparison}. The performance of the models improve consistently with the increasing hidden dimension size. This shows that learning capacity of the model benefits from more dimensions. The model with \texttt{circuit 10} outperforms the rest consistently. However, there seems to be a saturation of the best loss as the hidden dimension size increases. Training curves of these comparisons are presented in Appendix~\ref{appendix:dim}.

\begin{figure}
    \centering
    \includegraphics[width=\linewidth]{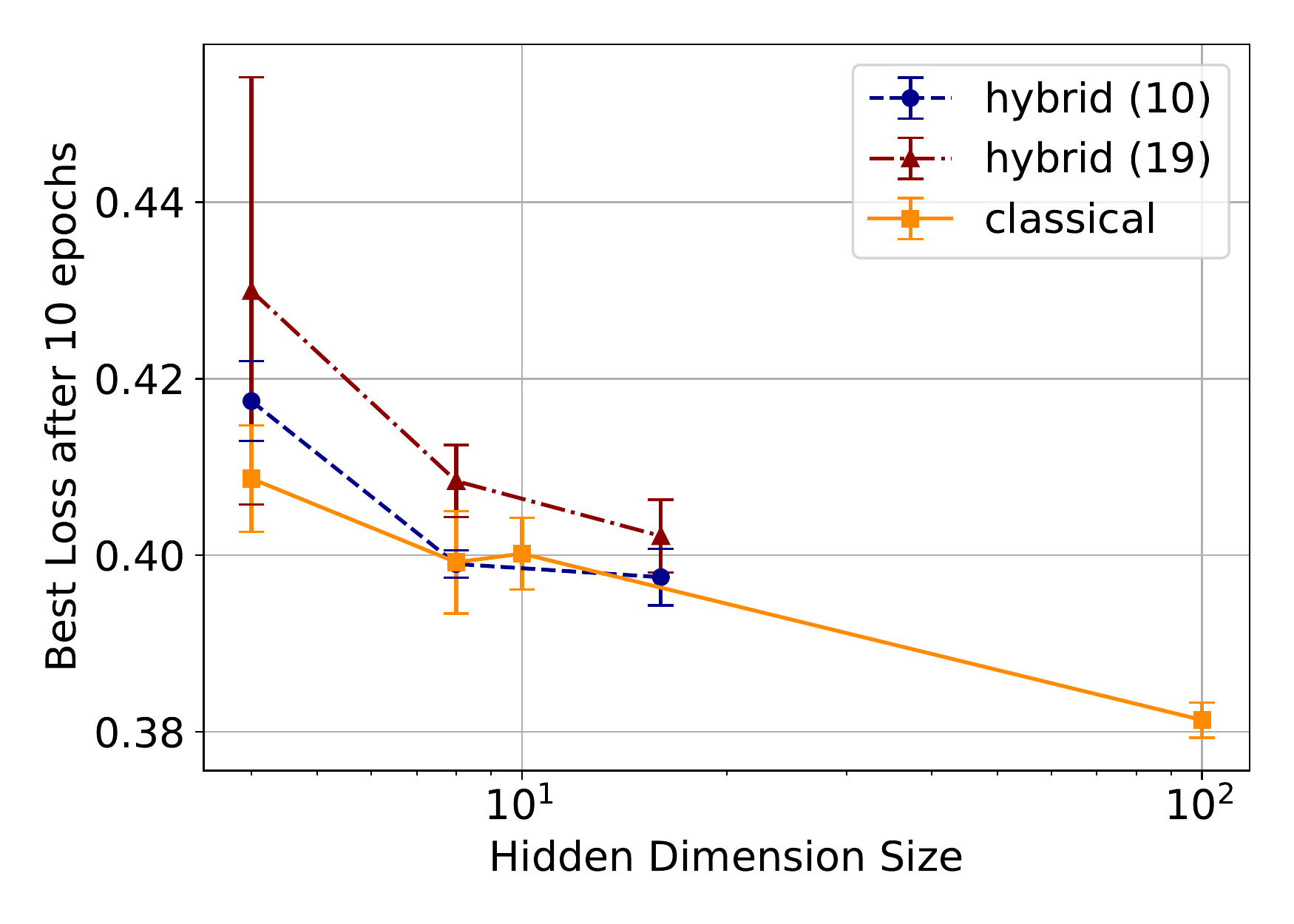}
    \caption{Best validation loss comparison with respect to different hidden dimension sizes ($N_D$) of the hybrid and classical GNN models. The comparison is made with the choice of $N_Q=N_D$, $N_I=3$ and $N_L=1$. 5 instances of all models with different initial parameters are trained for 10 epochs, and the mean of best losses are presented. The error bars represent the $\pm$ standard deviation of the best losses of all 5 runs.}
    \label{fig:classical_comparison}
\end{figure}

Finally, the hybrid model is compared against the classical model at different hidden dimension sizes and presented in Fig.~\ref{fig:classical_comparison}. For this comparison the same choice of $N_Q=N_D$, $N_I=3$ and $N_L=1$ is followed. This result shows that the hybrid model scales similarly to the classical model until a certain hidden dimension size. We did not perform simulations for qubits larger than 16 due to restrictions set by simulation times and classical hardware resources. 

\section{Conclusion}
\label{conclusion}

In this work, we implemented a hybrid quantum-classical GNN (QGNN) model for particle track reconstruction using the TrackML dataset(~\cite{trackml}). This is the first end-to-end implementation of a hybrid quantum-classical GNN model with attention passing to the best of our knowledge. We showed that the model can perform similar to classical approaches for low number of hidden dimension sizes. We investigated how the model scales for different hyper-parameters. \texttt{circuit 10} consistently performed the best among other models in all comparisons. On the other hand, \texttt{circuit 10} has the worst expressibility and is the lowest entangling model in a single layer configuration. Numerical results indicate that larger PQC models are harder to train, as it was shown in many instances (\cite{barren, Leyton-Ortega2021}). 

The current status of Quantum hardware restricted us to use only simulations of Quantum Circuits. This was mainly due to thousands of circuit executions required by the model. Because of the high pile-up conditions of the TrackML dataset, the graphs have thousands of nodes and edges, and therefore using hardware is a challenge for this approach. A forward pass of the presented QGNN model builds $N_I + 1$ circuits for each edge and $N_I$ circuits for each node. This also limited us experimenting with larger sized models due to restrictions in simulating Quantum Circuits. In order to cope with this problem, we enforced a $pT$-cut for reducing the number of particles, a small number of qubits (up to 16) was used. We also used analytical results with no noise and trained models up to 20 epochs at maximum. Very large RAM requirements and the significant increase of training times to more than a week for models with 16 qubits were the limiting factor in our results. 

This work explores an advantage in reducing the size of high dimensional NN layers with Quantum Circuits that have significantly less qubits. However, results obtained with $N_Q=N_D$ were only able to match the performance of the classical model. On the other hand, this does not mean that an advantage is still not possible. There is more to explore to better understand the potential of this approach.

First, the QGNN model was only experimented with simple encoding circuits (angle encoding), while more sophisticated encodings are conjectured to significantly affect the performance of QNN based models~(\cite{schuld_effect_2021}). Furthermore, recent work by~\cite{abbas_power_2020} showed that QNN models with four qubits and ZZ Feature Maps of depth two has a larger effective dimension compared to its classical equivalent, which leads to a better learning capacity. Therefore, a further study is needed to explore the potential benefits of different data encodings.

On top of that, this work does not explore any noise effects, which is considered as one of the limiting factors of VQAs as it can lead to Barren Plateaus~(\cite{wang2021noiseinduced}). Hardware noise is conjectured to slow down training of VQAs. However, this does not mean that noise always disfavors VQAs. It is argued that hardware noise can help explore the loss landscape~\cite{cerezo_variational_2021}. In many instances, VQAs were shown to have noise resilience~(\cite{sharma_noise_2020,gentini_noise-resilient_2020}) and benefit from noise~\cite{cao_noise-assisted_2021, campos_training_2021}. Furthermore \cite{mari_estimating_2021} showed that gradients and higher order of derivatives can be accurately obtained under hardware and shot noise. The results from the literature indicate that understanding the effect of noise on variational models is essential to estimate their potential. In this work, experiments with noise were attempted but then abandoned due to technical limitations posed by the size of the dataset.

The QGNN model can be further improved by employing better training schemes (\cite{Leyton-Ortega2021}) and noise aware optimizers~\cite{arrasmith2020operator}. Further research directions include exploring more sophisticated data encodings and understanding effect of noise. It would be more beneficial to work with a smaller dataset for exploring hybrid GNN models that target NISQ hardware.
\begin{acknowledgements}
Authors would like to thank Alessandro Roggero for fruitful discussions. Part of this work was conducted at "\textit{iBanks}", the AI GPU cluster at Caltech. We acknowledge NVIDIA, SuperMicro and the Kavli Foundation for their support of "\textit{iBanks}". This work was partially supported by Turkish Atomic Energy Authority (TAEK) (Grant No: 2017TAEKCERN-A5.H6.F2.15 and 2020TAEK(CERN)-A5.H1.F5-26).
\end{acknowledgements}
\section*{Software Information}
The open-source software used in this work can be listed as follows. Python 3.8.5, NumPy v1.18.5 (\cite{numpy}), Cirq v0.9.1 (\cite{cirq}), Tensorflow  v2.3.1 (\cite{abadi2016tensorflow}), Tensorflow Quantum v0.4.0 (\cite{tfq}), Scikit-learn v0.23.2 (\cite{scikit-learn}), qpic v1.0.2 (\cite{qpic}), Matplotlib v3.2.2 (\cite{matplotlib}). The project codebase to reproduce all of the results presented here can be accessed through \url{https://qtrkx.github.io}.
\section*{Conflict of interest}
The authors declare that they have no conflict of interest.
\bibliographystyle{spbasic}      
\bibliography{main}   
\clearpage
\onecolumn
\appendix
\section{The Dataset and Pre-processing Details}
\label{appendix:preprocessing}
\begin{figure}[!ht]
    \centering
    \includegraphics[width=.6\linewidth]{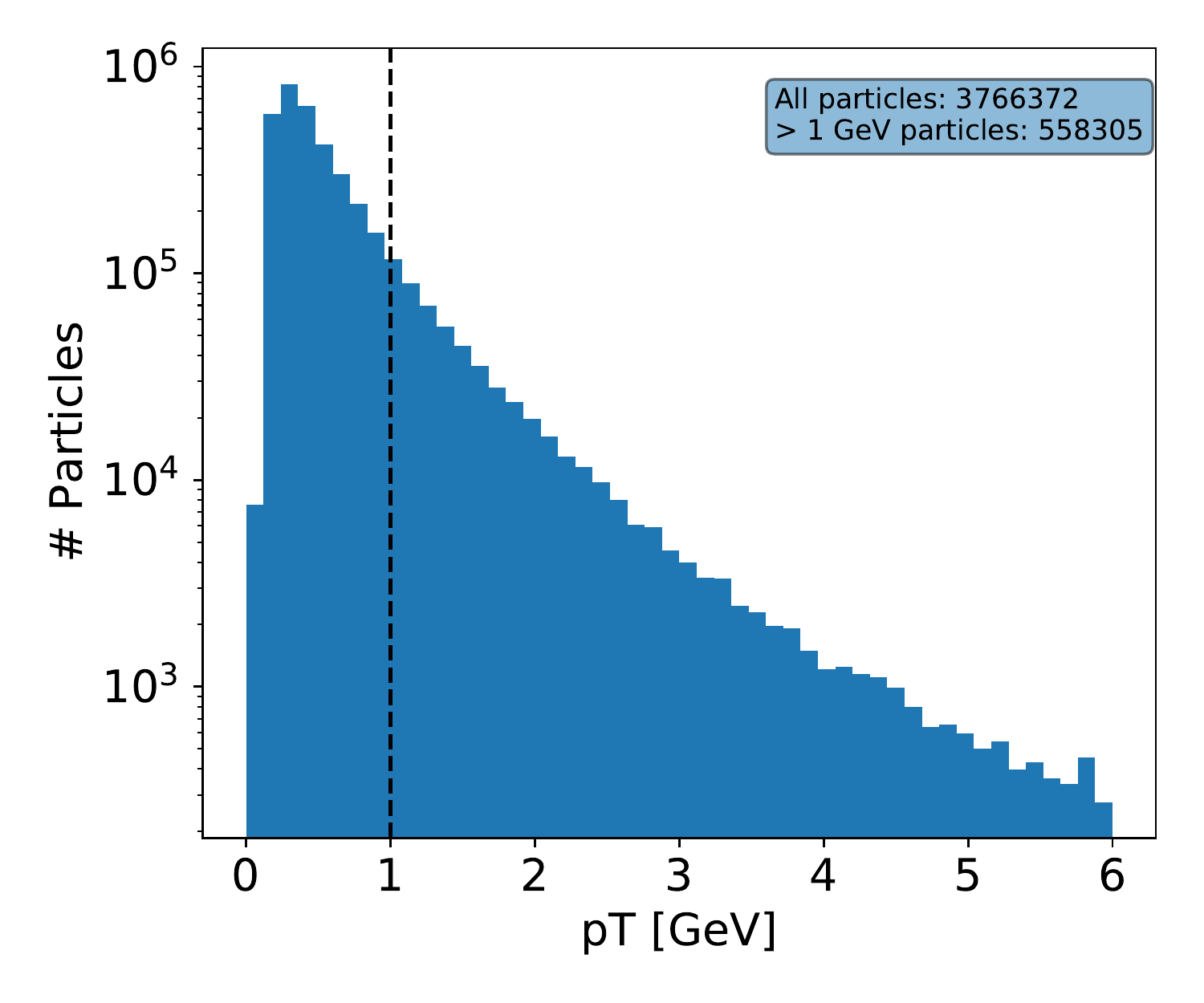}
    \caption{Histogram of number of particles in 100 events that are inside the barrel region of the TrackML detector. The dashed line represents the 1 GeV pT threshold we have selected in order to reduce total number of particles and tracks.}
\end{figure}
\begin{figure}[!ht]
    \centering
    \includegraphics[width=\linewidth]{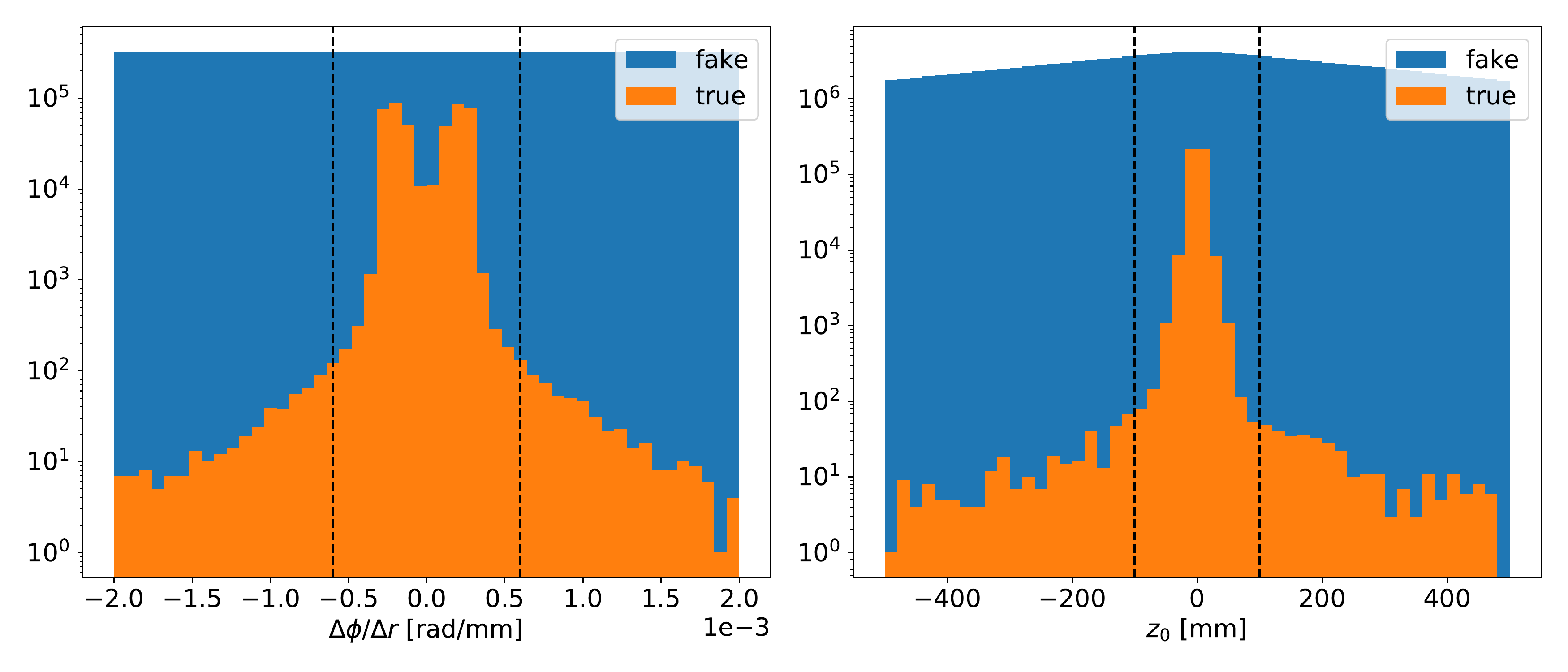}
    \caption{Histogram of fake and true segments obtained during the graph construction procedure. The segments that reside inside the dashed lines are considered in order to maximize purity and efficiency. This step is important as it allows graph construction with fewer fake edges, and thus reduces computation times significantly.}
\end{figure}
\clearpage
\section{Expressibility and Entanglement Capability of PQCs}
\label{appendix:exp-ent}
\begin{figure*}[!ht]
    \centering
    \includegraphics[width=.6\textwidth]{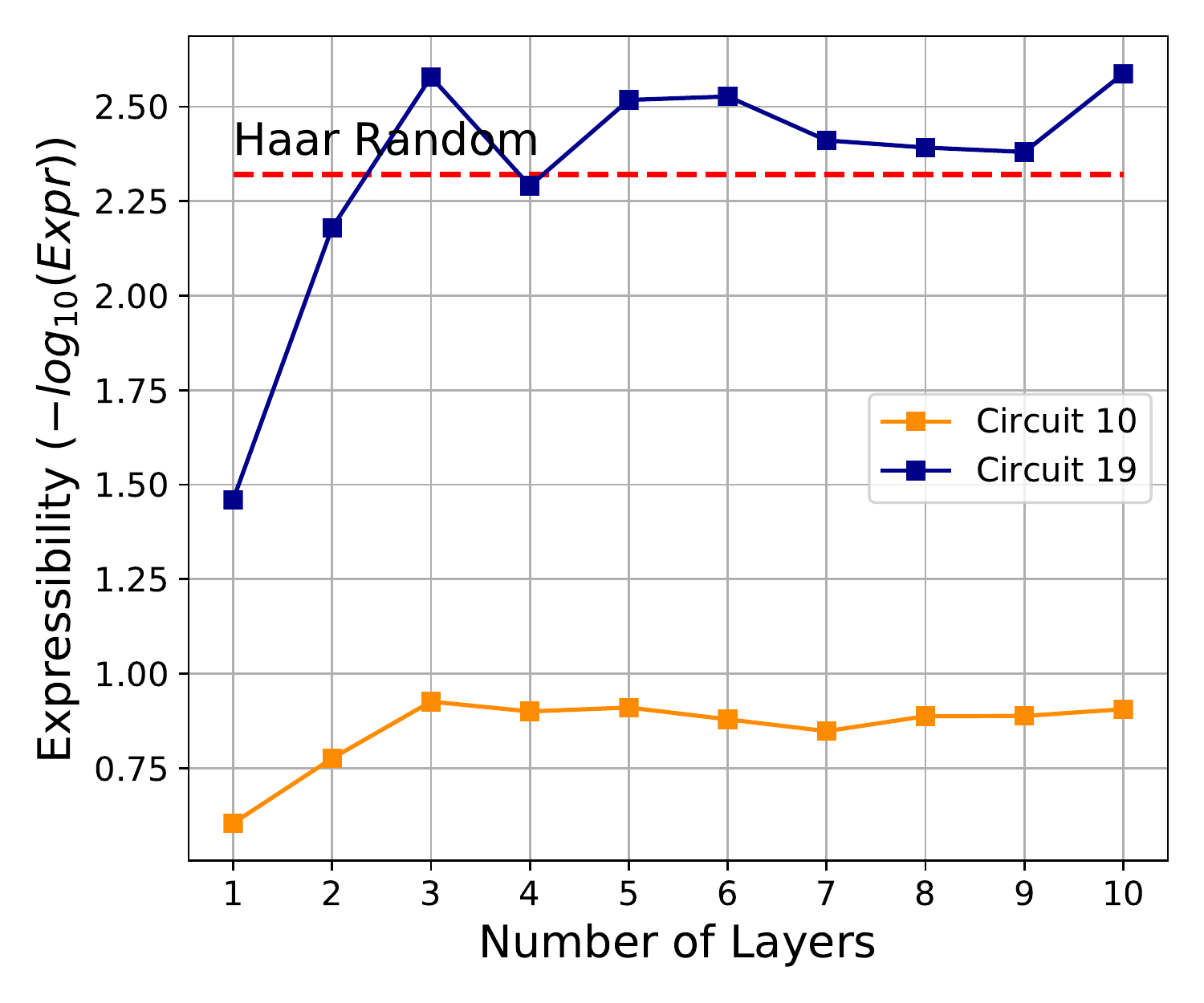}
    \caption{Expressibility vs. number of layers for Circuit 10 (orange) and Circuit 19 (blue) in their 4 qubit configurations. Negative log of Expressibility is plotted for visual purposes.}
    \label{fig:exp}
\end{figure*}

\begin{figure*}[!ht]
    \centering
    \includegraphics[width=.6\textwidth]{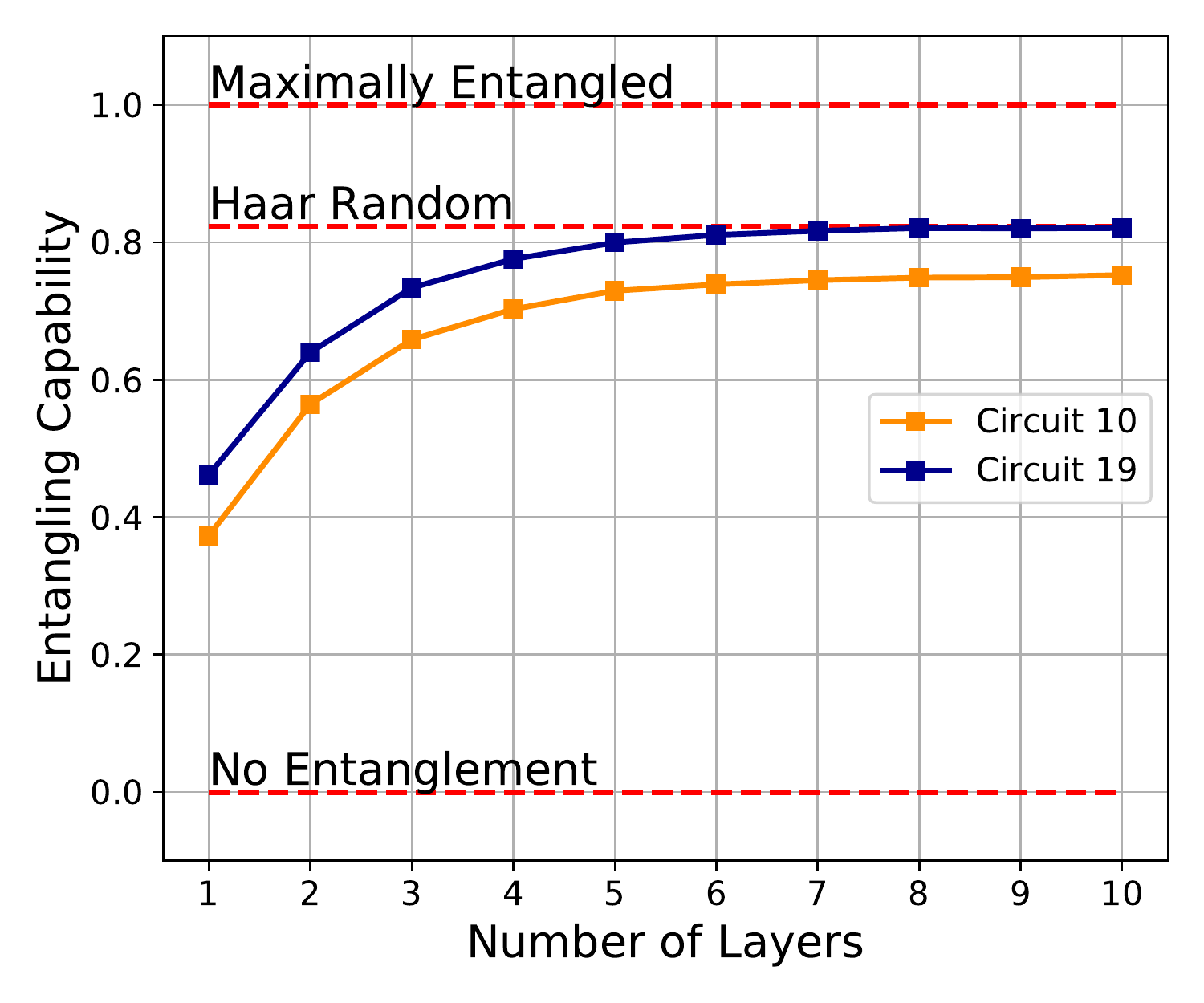}
    \caption{Entanglement Capability vs. number of layers for Circuit 10 (orange) and Circuit 19 (blue) in their 4 qubit configurations.}
    \label{fig:ent}
\end{figure*}
\clearpage
\section{Hidden Dimension Size Comparison}
\label{appendix:dim}
\begin{figure}[!ht]
    \centering
    \includegraphics[width=\linewidth]{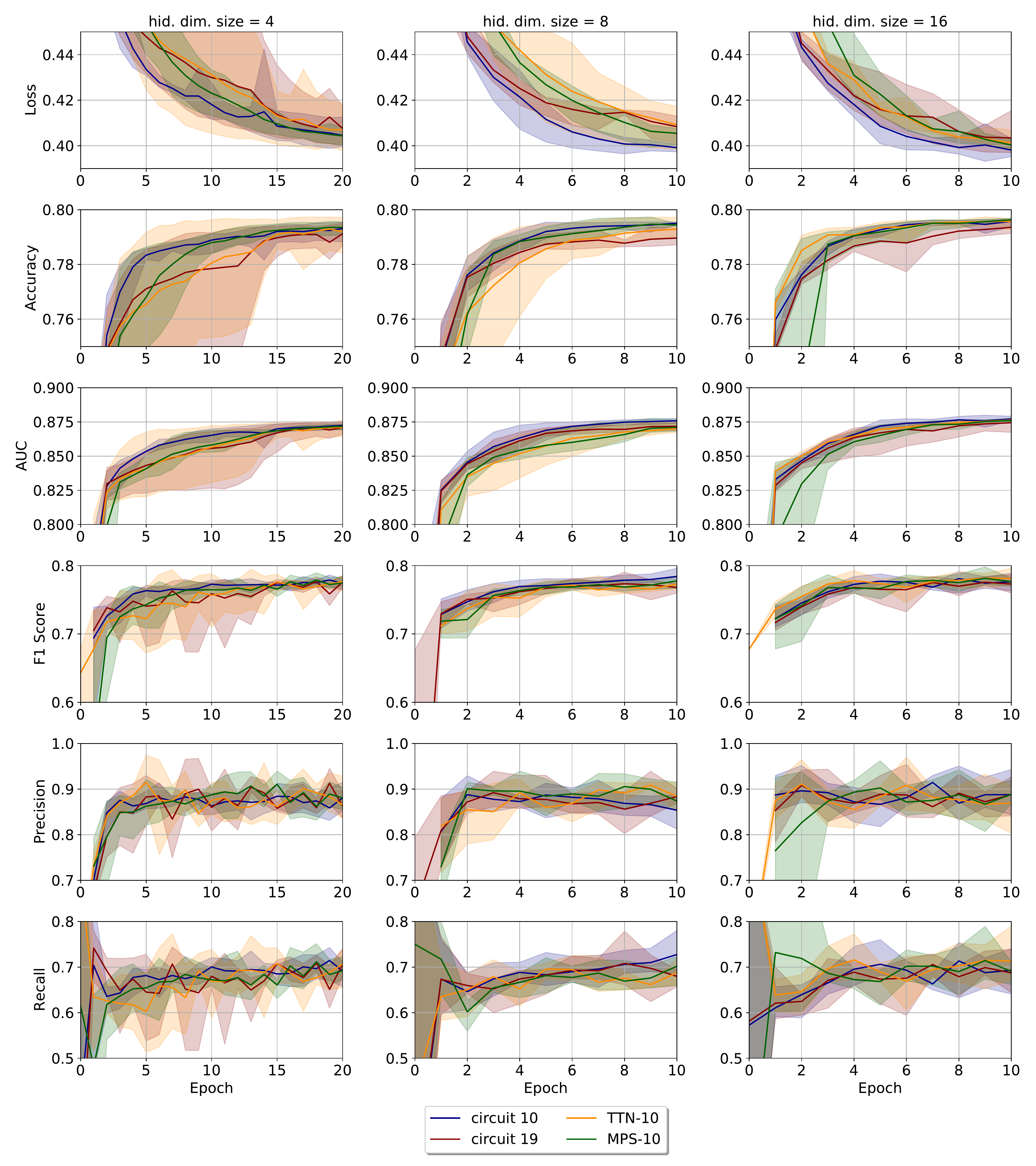}
    \caption{Hidden dimension size comparison compares different hidden dimension sizes ($N_D$) by fixing $N_Q=N_D$, $N_I=3$ and $N_L=1$. The shaded regions represent the $\pm$ standard deviation of the best losses of all 5 runs. A 0.5 threshold is used for metrics that require a threshold.}
\end{figure}
\clearpage
\section{Number of Layers Comparison}
\label{appendix:layer}
\begin{figure}[!ht]
    \centering
    \includegraphics[width=\linewidth]{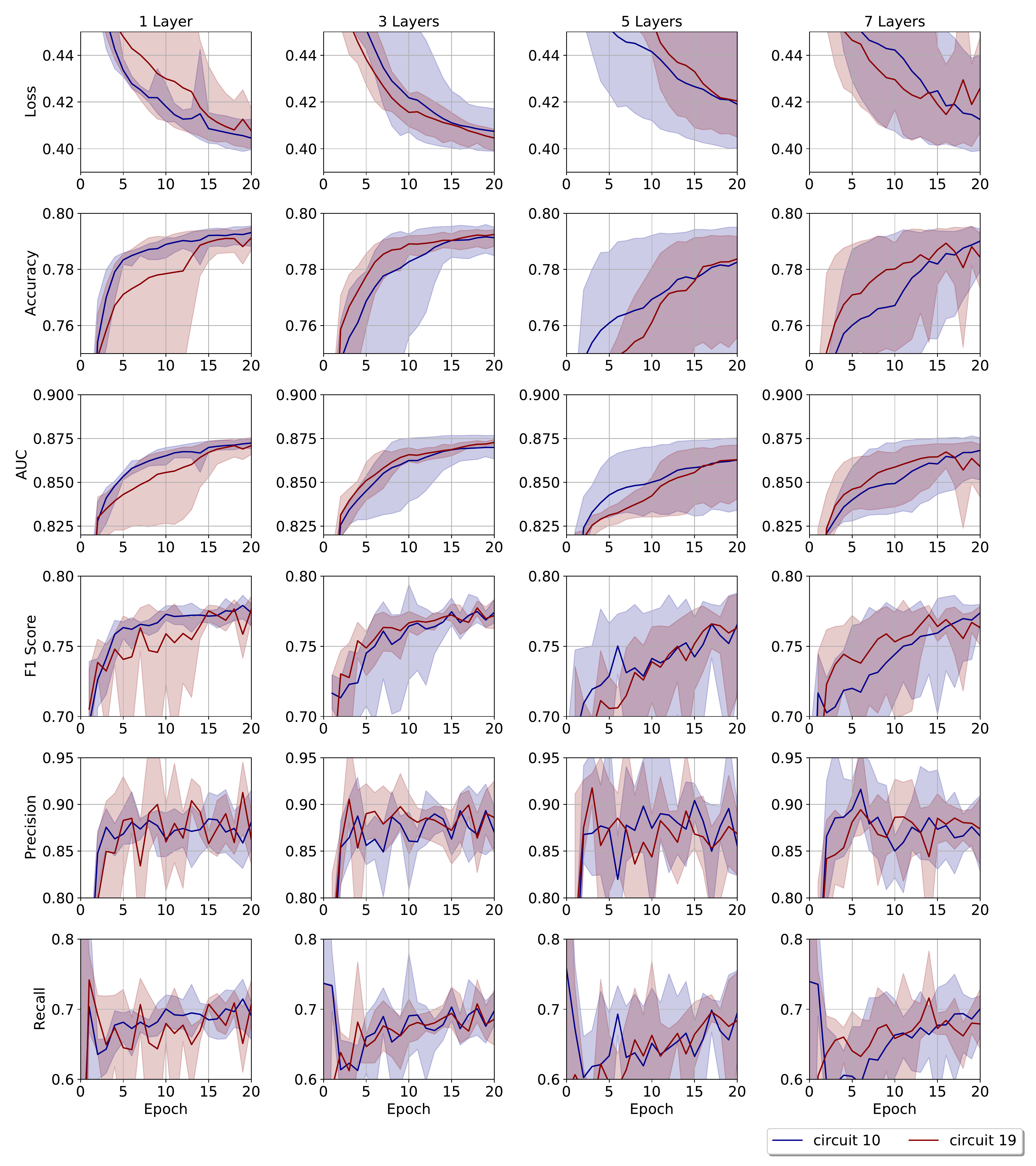}
    \caption{Number of layers comparison compares number of layers by fixing $N_Q=N_D=4$ and $N_I=3$. The shaded regions represent the $\pm$ standard deviation of the best losses of all 5 runs. A 0.5 threshold is used for metrics that require a threshold.}
\end{figure}
\clearpage
\section{Number of Iteration Comparison}
\label{appendix:iter}
\begin{figure}[!ht]
    \centering
    \includegraphics[width=\linewidth]{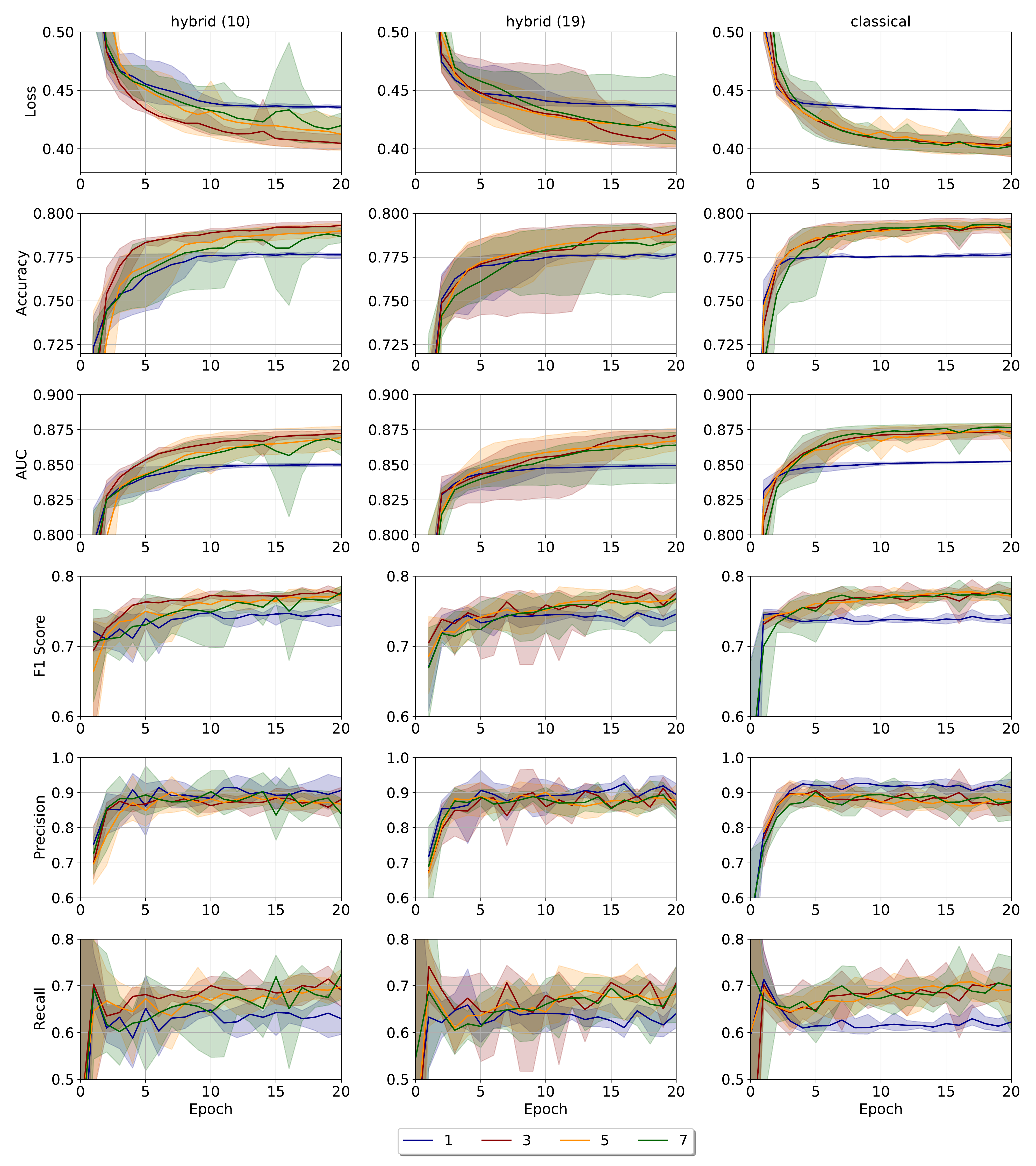}
    \caption{Number of iteration comparison compares number of iteration by fixing $N_Q=N_D=4$ and $N_L=1$. The shaded regions represent the $\pm$ standard deviation of the best losses of all 5 runs. A 0.5 threshold is used for metrics that require a threshold.}
\end{figure}
\clearpage
\section{Embedding Comparison}
\label{appendix:embedding}
\begin{figure}[!ht]
    \centering
    \includegraphics[width=\linewidth]{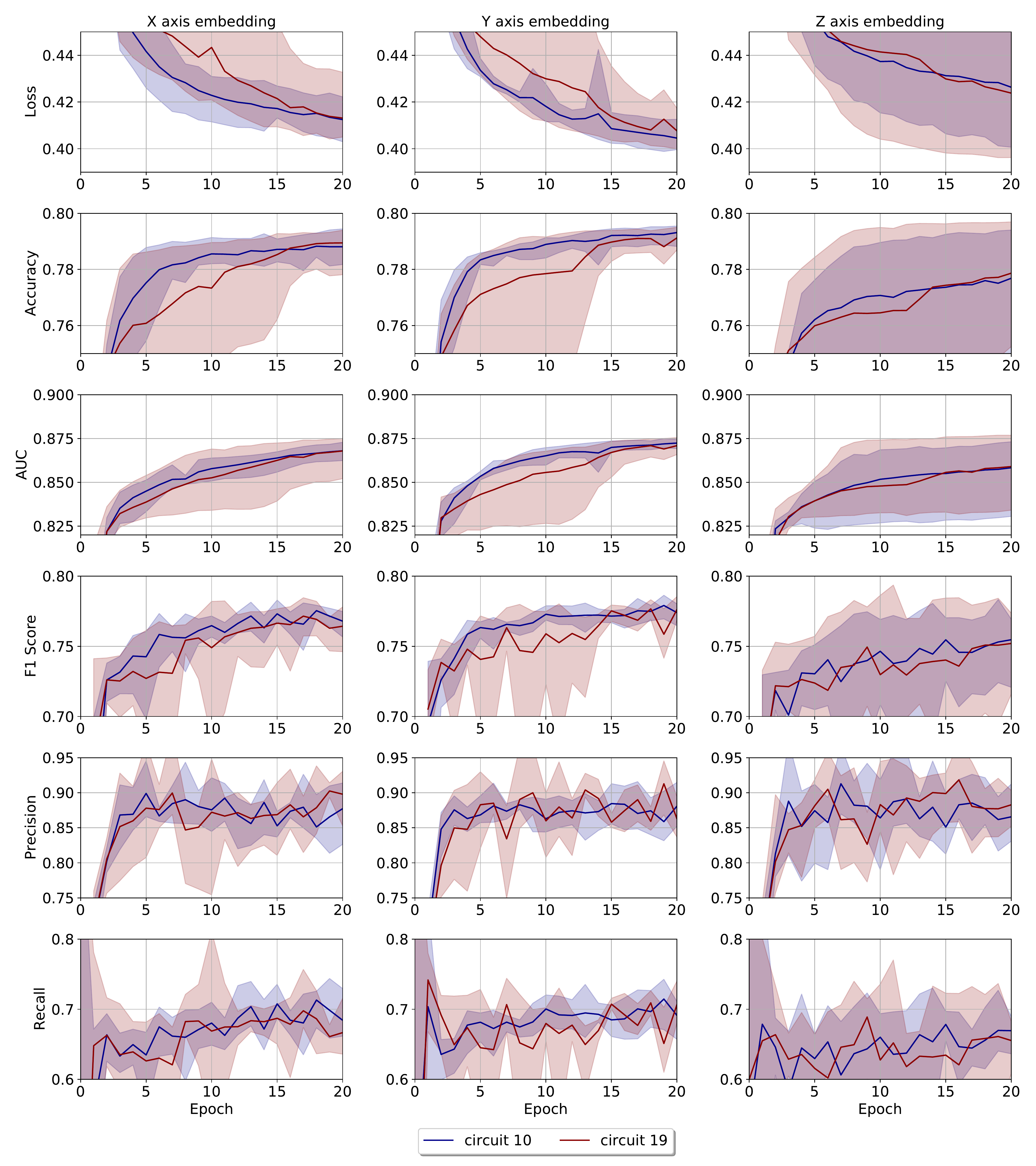}
    \caption{Axis of angle embedding comparison compares different embedding axes by fixing $N_Q=N_D=4$, $N_I=3$ and $N_L=1$. The shaded regions represent the $\pm$ standard deviation of the best losses of all 5 runs. A 0.5 threshold is used for metrics that require a threshold.}
\end{figure}
\end{document}